\documentclass[aps,prb,twocolumn,superscriptaddress,showpacs]{revtex4-1}
\usepackage{amsmath,amssymb,amsfonts,amsthm}
\usepackage{graphicx}
\usepackage{bm}
\usepackage{bbm}
\usepackage{color}
\usepackage{dcolumn}   
\usepackage{epstopdf}
\usepackage[caption=false]{subfig}
\usepackage{color}
\usepackage[colorlinks=true,linkcolor=blue,urlcolor=blue,citecolor=blue]{hyperref}
\usepackage{tikz}
\usepackage{eucal}
\usepackage[top=1in,bottom=1in,left=1in,right=1in]{geometry}
\usepackage[normalem]{ulem}

\begin{document}

 \newcommand{\breite}{1.0} 

\newtheorem{prop}{Proposition}
\newtheorem{cor}{Corollary} 

\newcommand{\be}{\begin{equation}}
\newcommand{\ee}{\end{equation}}

\newcommand{\bea}{\begin{eqnarray}}
\newcommand{\eea}{\end{eqnarray}}
\newcommand{\lt}{<}
\newcommand{\gt}{>}

\newcommand{\Reals}{\mathbb{R}}     
\newcommand{\Com}{\mathbb{C}}       
\newcommand{\Nat}{\mathbb{N}}       
\def\e{\epsilon}

\newcommand{\id}{\mathbboldsymbol{1}}    

\newcommand{\Real}{\mathop{\mathrm{Re}}}
\newcommand{\Imag}{\mathop{\mathrm{Im}}}

\def\O{\mbox{$\mathcal{O}$}}   
\def\F{\mathcal{F}}			
\def\sgn{\text{sgn}}

\newcommand{\deo}{\ensuremath{\Delta_0}}
\newcommand{\dea}{\ensuremath{\Delta}}
\newcommand{\ak}{\ensuremath{a_k}}
\newcommand{\ad}{\ensuremath{a^{\dagger}_{-k}}}
\newcommand{\sx}{\ensuremath{\sigma_x}}
\newcommand{\sz}{\ensuremath{\sigma_z}}
\newcommand{\spl}{\ensuremath{\sigma_{+}}}
\newcommand{\smi}{\ensuremath{\sigma_{-}}}
\newcommand{\alk}{\ensuremath{\alpha_{k}}}
\newcommand{\bk}{\ensuremath{\beta_{k}}}
\newcommand{\ok}{\ensuremath{\omega_{k}}}
\newcommand{\vd}{\ensuremath{V^{\dagger}_1}}
\newcommand{\vi}{\ensuremath{V_1}}
\newcommand{\vo}{\ensuremath{V_o}}
\newcommand{\zc}{\ensuremath{\frac{E_z}{E}}}
\newcommand{\xc}{\ensuremath{\frac{\Delta}{E}}}
\newcommand{\xd}{\ensuremath{X^{\dagger}}}
\newcommand{\aok}{\ensuremath{\frac{\alk}{\ok}}}
\newcommand{\tpw}{\ensuremath{e^{i \ok s }}}
\newcommand{\tpe}{\ensuremath{e^{2iE s }}}
\newcommand{\tmw}{\ensuremath{e^{-i \ok s }}}
\newcommand{\tme}{\ensuremath{e^{-2iE s }}}
\newcommand{\epls}{\ensuremath{e^{F(s)}}}
\newcommand{\emis}{\ensuremath{e^{-F(s)}}}
\newcommand{\epl}{\ensuremath{e^{F(0)}}}
\newcommand{\emi}{\ensuremath{e^{F(0)}}}

\newcommand{\p}{\partial}
\newcommand{\lr}[1]{\left( #1 \right)}
\newcommand{\lrs}[1]{\left( #1 \right)^2}
\newcommand{\lrb}[1]{\left< #1\right>}
\newcommand{\nbt}{\ensuremath{\lr{ \lr{n_k + 1} \tmw + n_k \tpw  }}}

\newcommand{\om}{\ensuremath{\omega}}
\newcommand{\dw}{\ensuremath{\Delta_0}}
\newcommand{\wbp}{\ensuremath{\omega_0}}
\newcommand{\dv}{\ensuremath{\Delta_0}}
\newcommand{\vbp}{\ensuremath{\nu_0}}
\newcommand{\vplus}{\ensuremath{\nu_{+}}}
\newcommand{\vminus}{\ensuremath{\nu_{-}}}
\newcommand{\wplus}{\ensuremath{\omega_{+}}}
\newcommand{\wminus}{\ensuremath{\omega_{-}}}
\newcommand{\uv}[1]{\ensuremath{\mathbf{\hat{#1}}}} 
\newcommand{\abs}[1]{\left| #1 \right|} 
\newcommand{\avg}[1]{\langle #1 \rangle} 
\let\underdot=\d 
\renewcommand{\d}[2]{\frac{d #1}{d #2}} 
\newcommand{\dd}[2]{\frac{d^2 #1}{d #2^2}} 
\newcommand{\pd}[2]{\frac{\partial #1}{\partial #2}} 
\newcommand{\pdd}[2]{\frac{\partial^2 #1}{\partial #2^2}} 
\newcommand{\pdc}[3]{\left( \frac{\partial #1}{\partial #2}
 \right)_{#3}} 
\newcommand{\ket}[1]{\left| #1 \right>} 
\newcommand{\bra}[1]{\left< #1 \right|} 
\newcommand{\braket}[2]{\left< #1 \vphantom{#2} \right|
 \left. #2 \vphantom{#1} \right>} 
\newcommand{\matrixel}[3]{\left< #1 \vphantom{#2#3} \right|
 #2 \left| #3 \vphantom{#1#2} \right>} 
\newcommand{\grad}[1]{{\nabla} {#1}} 
\let\divsymb=\div 
\renewcommand{\div}[1]{{\nabla} \cdot \boldsymbol{#1}} 
\newcommand{\curl}[1]{{\nabla} \times \boldsymbol{#1}} 
\newcommand{\laplace}[1]{\nabla^2 \boldsymbol{#1}}
\newcommand{\vs}[1]{\boldsymbol{#1}}
\let\baraccent=\= 
\newcommand{\red}[1]{{\color{red}#1}}
\newcommand{\blue}[1]{{\color{blue}#1}}

\title{Cooling arbitrary near-critical systems using hyperbolic quenches}

\author{Prahar Mitra}
\affiliation{School of Natural Sciences, Institute for Advanced Study, Princeton, NJ 08540, USA}
\author{Matteo Ippoliti}
\affiliation{Department of Physics, Princeton University, Princeton, New Jersey 08544, USA}
\author{R. N. Bhatt}
\affiliation{Department of Electrical Engineering, Princeton University, Princeton, New Jersey 08540, USA}
\author{S. L. Sondhi}
\affiliation{Department of Physics, Princeton University, Princeton, New Jersey 08544, USA}
\author{Kartiek Agarwal} 
\email{kagarwal@princeton.edu}
\affiliation{Department of Electrical Engineering, Princeton University, Princeton, New Jersey 08540, USA}

\date{\today}
\begin{abstract}
We describe a quench protocol that allows the rapid preparation of ground states of arbitrary interacting conformal field theories in $1+1$ dimensions. We start from the ground state of a related gapped relativistic quantum field theory and consider sudden quenches along the space-like trajectories $t^2 - x^2 = T^2_0$ (parameterized by $T_0$) to a conformal field theory. Using only arguments of symmetry and conformal invariance, we show that the post-quench stress-energy tensor of the conformal field theory is uniquely constrained up to an overall scaling factor. Crucially, the \emph{geometry} of the quench necessitates that the system approach the vacuum energy density over all space except the singular lines $x = \pm t$. The above arguments are verified using an exact treatment of the quench for the Gaussian scalar field theory (equivalently the Luttinger liquid), and numerically for the quantum $O(N)$ model in the large-$N$ limit. Additionally, for the Gaussian theory, we find in fact that even when starting from certain excited states, the quench conserves entropy, and is thus also suitable for rapidly preparing excited states. Our methods serve as a fast, alternative route to reservoir-based cooling to prepare quantum states of interest.   
\end{abstract}
 
\maketitle

\section{Introduction.}
 
Experimentally engineering and harnessing the power of artificial quantum systems for the purpose of quantum simulation and quantum computation is an important present challenge. While much progress has been made on the front of developing extremely isolated quantum systems---ultracold atoms in optical lattices~\cite{mazurenko2017cold,cheuk2016observation,schneiderMBL,luschen2016evidence} or traps~\cite{langen2015ultracold,Langen}, nitrogen vacancy centers~\cite{fuchsNV,DuttNV,MaurerNV,ShuskovNV,AgarwalmagneticnoiseNV,choi2017observation}, ion traps~\cite{debnath2016demonstration,zhang2017observation,jurcevic2014observation}, superconducting qubit structures~\cite{barends2016digitized,boixo2013quantum,johnson2011quantum,dwaveannealing} etc.---as these systems grow more complex, it becomes harder to devise equally elaborate tools to manipulate them while maintaining isolation from sources of decoherence. It is therefore important to theoretically determine the minimum set of control knobs needed to prepare certain quantum states of interest, and the most efficient way to do so. This is the challenge of quantum state preparation. 

In this regard, the adiabatic principle has served as a basis for many investigations (cf. Ref.~\cite{griffiths1995introduction}). In its simplest form, the idea is to prepare the system in an eigenstate of a Hamiltonian that is easily accessible---usually gapped, such that the ground state lacks long-range entanglement---and subsequently tune the Hamiltonian slowly to evolve this eigenstate into the target state. When this action is performed sufficiently slowly, the system continues to evolve in an eigenstate of the instantaneous Hamiltonian. The limitation of this approach is its speed---to avoid exciting the system in the process, the time taken must be of the order of the inverse-square of the smallest instantaneous spectral gap between the target and excited states~\cite{mingapYoung}, a quantity which diverges in the thermodynamic limit for many systems/problems of interest~\cite{QAAsatisfiability,QAAparticular,AgarwalGroundStatePrep}. 

To achieve faster preparation, recent work has proposed engineering counter-diabatic drives~\cite{delCampoCounterDiabatic,del2013shortcuts,jarzynskitransitionless,glaser1998unitary,sels2016minimizing} that counter the production of excitations during adiabatic evolution, or introducing `optimal-control' protocols~\cite{van2016optimal,Jiangfeng2016optimalspinqubit,Superadiabaticspinchaintransfer2017,aashishcleark_diabatic,bulkspinoptimalcontrol,ho2018efficient} (including `bang-bang' protocols~\cite{speedlimitHegerfeldt,bao2017optimal,yang2017optimizing,bukov2017machine}) that entirely dispense with the adiabatic ansatz. While these methods indeed outpace adiabatic protocols, they often rely on extensive numerical simulations to explore the parameter space of preparation protocols to find the optimal one; importantly insights from protocols found for finite size systems do not appear to carry over in an obvious way to the thermodynamic limit. For present experimentally achievable system sizes~\cite{mazurenko2017cold}, it is still most efficient to create a thermodynamically large reservoir of low-energy excitations~\cite{ho2009universal,zaletel2016preparation} that can remove entropy from the subsystem of interest; this may however prove challenging to extend to larger systems, systems that exhibit integrability, and systems that themselves have low-energy excitations.   

\begin{center}
\begin{figure}
\includegraphics[width = 3.1in]{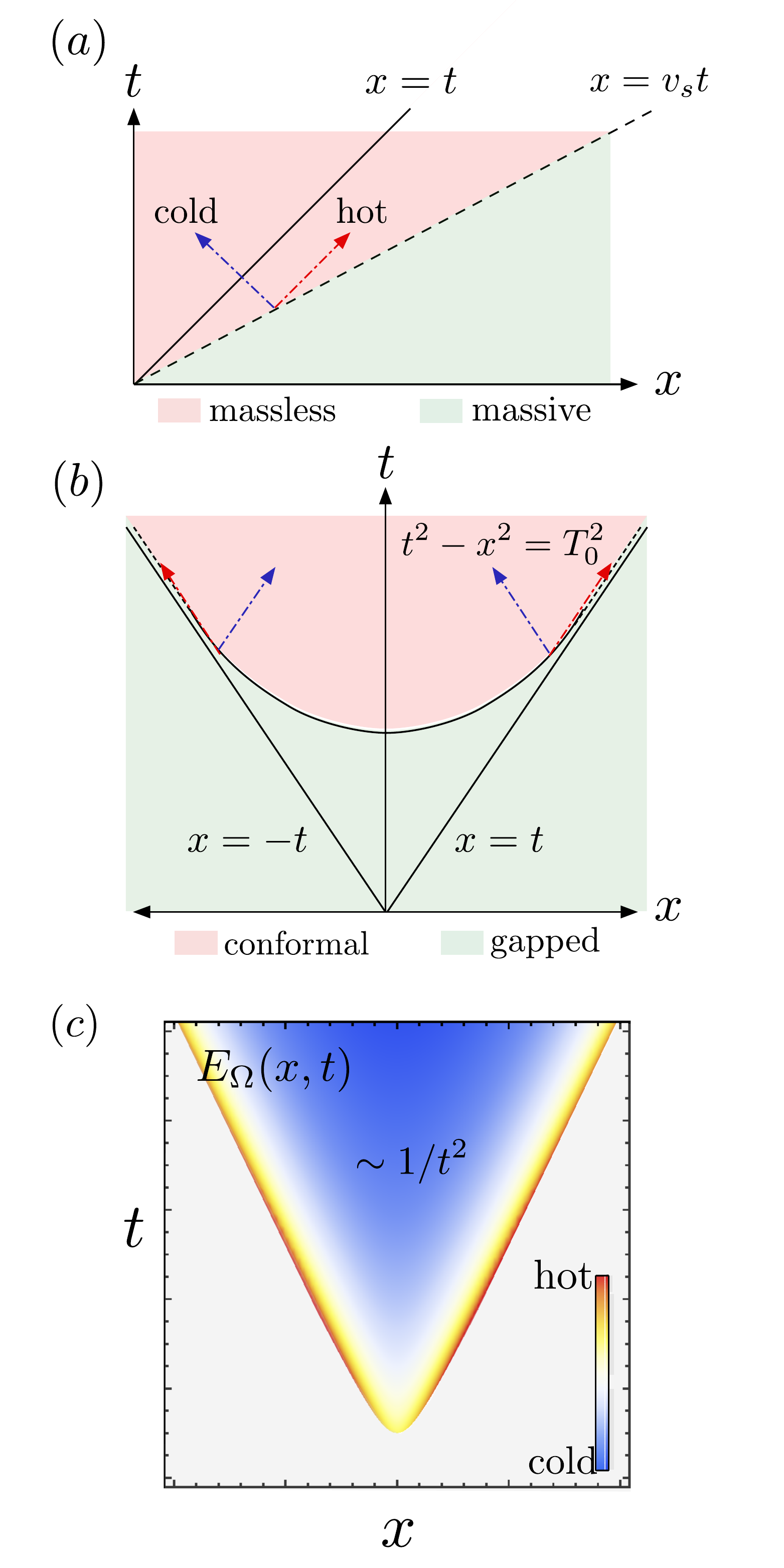}
\caption{(a) Protocol studied in Ref.~\cite{AgarwalGroundStatePrep}. (b) Hyperbolic quench protocols considered in this work. (c) Schematic representation of the spatiotemporal dependence of the post-quench energy-density as fixed by conformal symmetries, in the post-quench region $t^2 - x^2 \ge T^2_0$. ``Heat waves" emerge~\cite{Agarwalquantumheatwaves} near the boundaries of the quench trajectory, but become localized near $x = \pm t$ over time. The energy density decays as $\sim 1/t^2$ everywhere else, approaching the minimum $\sim 1/L^2$ in time $\sim \mathcal{O}[L]$.}
\label{fig:figone}
\end{figure}
\end{center}

In this work, we build upon previous work~\cite{AgarwalGroundStatePrep,Agarwalquantumheatwaves} (and also related work in the Kibble-Zurek community, see Refs.~\cite{Dziarmagainhomogeneous,Dziarmagazgreaterthanone}) by some of the present authors and discuss a general paradigm for preparing the ground state of \emph{arbitrary interacting} conformal field theories (CFTs) in spatial dimension $d=1$. Such systems are particularly challenging to cool because they harbor gapless excitations, and are often integrable (e.g., Luttinger liquids). As in previous work, we assume that initially the system resides in the ground state of a related gapped relativistic quantum field theory (QFT) which is easier to prepare due to the presence of a gap. We next consider a quench to the CFT of interest by eliminating the relevant perturbation that gaps out the low-energy modes in the QFT along a special space-time trajectory. While the previous work considered a quench wherein the local mass was set to zero along a superluminally moving front, here we consider a hyperbolic quench trajectory, $(ct)^2 - x^2 = T^2_0$; the two protocols are illustrated in Figure \ref{fig:figone}. As we show, this class of quench trajectories represents an entire family of ground-state preparation protocols (delineated by $T_0$) that prepare the ground state in time $t \sim \mathcal{O} \left[ L \right]$ where $L$ is the system size; the optimal luminal quench considered previously corresponds to the limit $T_0 \rightarrow 0$. More importantly, studying this family of quenches allows us to uncover the purely \emph{geometric} origins of the cooling mechanism of the luminal quenches of Ref.~\cite{AgarwalGroundStatePrep}, at least in $d = 1$, thus proving rigorously their use for cooling arbitrary interacting CFTs.

More concretely, it was previously argued that the superluminal motion of the quench front resulted in the production of a chiral population of excitations~\cite{AgarwalChiral,Agarwalquantumheatwaves,AgarwalGroundStatePrep}. In the case of the optimal luminal quench, all excitations moving against the front were found to be Doppler-shifted to zero energy, while the excitations moving along were infinitely excited. These hot excitations pile up (in a way similar to a sonic boom) at the quench front. As a result, all the dissipation in the quench protocol is swept away in an infinitesimally sharp front, allowing the rapid preparation of the vacuum state everywhere else in the system. Generalizing such a protocol to an interacting setting where hot excitations can be reflected back is not obvious; however, plausible arguments and numerical data were provided to show how the introduction of a small amount of adiabaticity---by way of an additional time-scale for the local quenching of the gap---can make the protocol amenable to the interacting case.  

Here we show that the superluminal quench protocol considered previously is exact even for interacting systems by uncovering the geometric origins of the cooling process. In particular, the hyperbolic quenches considered here can be interpreted as occurring uniformly in space at a fixed time in \emph{conformal coordinates} $(\eta,\xi)$, defined by the relations $t = T_0 e^{\eta} \cosh \xi$ and $x = T_0 e^{\eta} \sinh \xi$. The quench involves the removal of the mass term at $\eta = 0$ for all $\xi$. Translations in $\xi$ correspond to an isometry of the system (equivalent to a Lorentz boost), the quench surface $\eta = 0$ as well as the initial pre-quench state (\emph{i.e.} the ground state of the QFT). As a result, the conformal system heats up uniformly in the conformal coordinates after the quench, and exhibits equilibrium in this coordinate system. In the laboratory frame, this appears to be a highly anisotropic non-equilibrium steady state. Crucially, in this state, the energy density approaches the vacuum energy as $\sim 1/t^2$ \emph{everywhere} except on the singular lines $x = \pm t$. Given the purely geometric foundations for the result, the protocol is applicable to arbitrary CFTs. (Note that our results bear resemblance to the observation of non-equilibrium steady states in critical systems related to stationary states in Lorentz-boosted frames~\cite{bhaseen2015energy,bernard2012energy,AgarwalChiral,castro2016emergent,lucas2016shock}; our system here looks static in a conformal coordinate system instead.) We show that the space and time dependence of the stress-energy tensor is completely constrained by arguments of symmetry and conformal invariance, and the only input specific to the quench protocol is a scaling function related to the initial energy density generated by the quench. Consistent with the previous work, we find that in the limit $T_0 \rightarrow 0$, that is, when the quench reduces to a symmetric copy of the luminal quench considered in Ref. \cite{AgarwalGroundStatePrep}, the energy density at all spacetime points away from the singular lines $t= \pm x$ is \emph{arbitrarily} close to that of the vacuum immediately away from the quench front. 

While more work needs to be done to appreciate the effectiveness of such hyperbolic quenches quenching from initially excited states, here we show that for the Gaussian theory at least, a uniform momentum-independent excited state mode population $n$ of massive bosons directly translates to the same population $n$ of massless bosonic excitations after the quench. In this way, the quench appears to preserve entropy (except along singular lines $x = \pm t$) more generally even when starting from these particular excited states. This result should be particularly useful for preparing ground states of one-dimensional systems described by a Luttinger liquid---if the temperature of the experimental system is initially below the mass gap, the excitations in the post-quench massless theory mirror the exponential suppression of the pre-quench massive theory due to the gap.  

This paper is organized as follows. In Sec.~\ref{sec:gened}, we discuss the formal argument for cooling in our protocol using symmetry and conformal invariance. In Sec.~\ref{sec:Gaussianquench}, we provide a validation of our findings by performing an exact calculation of the quench in a Gaussian scalar field theory. To provide additional verification in an interacting setting, in Sec.~\ref{sec:numerics} we examine our quench protocol in the quantum $O(N)$ model in the large-$N$ limit using numerical simulations. We complete the analysis in Sec.~\ref{sec:higherd} by showing that symmetry arguments alone do not constrain the stress-energy tensor in higher dimensions. We conclude by summarizing our results in Sec.~\ref{sec:conclusions}. 

\section{Argument for cooling}
\label{sec:gened}

We denote the laboratory coordinates as $x^a = (t,x)$ and define conformal coordinates $x^\mu = (\eta, \xi)$ which are related to the laboratory coordinates by 
\begin{equation}
\begin{split}
t = T_0 e^\eta \cosh \xi,  \qquad x = T_0 e^\eta \sinh \xi.
\end{split}
\end{equation}
Note that the conformal coordinates cover only the region $t \geq |x|$ of spacetime. The metric of Minkowski spacetime takes the form
\begin{equation}
\begin{split}\label{minkmetric}
ds^2 = - dt^2 + dx^2 = T_0^2 e^{2\eta} ( - d\eta^2 + d\xi^2 ) . 
\end{split}
\end{equation}

We assume that at $t = 0^-$, our system resides in the ground state of a particular gapped (gap $\sim m$) quantum field theory. We then argue that quenching the quantum field theory to a conformal field theory along the specific space-time trajectory $t^2 - x^2 = T^2_0$ allows one to approach the ground state of the conformal field theory rapidly.

Let us now understand the role of the conformal coordinates $(\eta, \xi)$. First, note that constant $\eta$ surfaces provide a hyperbolic foliation of the region $t > |x|$ of Minkowski spacetime into Cauchy surfaces defined by the relation $t^2 - x^2 = T^2_0 e^{2 \eta}$. The quench trajectory is a particular such Cauchy surface, defined by $\eta =0$. Further, we are interested in the description of our system within the post-quench region $t^2 - x^2 \ge T^2_0$ which is confined to the region $t \geq |x|$. For these reasons, it is clear that the conformal coordinates are more suitable for the problem at hand. To facilitate this, we must describe the initial pre-quench state at $\eta = 0^-$. This is done by a Hamiltonian evolution of the ground state at $t=0$ to $\eta = 0$ (choosing the appropriate notion of ``time'' which maps the $t=0$ Cauchy surface to the $\eta = 0$ one). Importantly, due to causality, one need only consider the pre-quench Hamiltonian for this evolution without any reference to the quench.

Another useful property of the conformal coordinates is that translation in $\xi$, that is $\xi \mapsto \xi+a$, is an isometry of the system, as is clear from the form of the Minkowski metric in conformal coordinates \eqref{minkmetric}. One can also see this by noting that a translation in $\xi$ corresponds to a Lorentz boost transformation in the laboratory frame coordinates: $(t,x) \mapsto (t \cosh a + x \sinh a , t \sinh a  + x \cosh a)$. This is therefore a symmetry of the pre-quench relativistic quantum field theory, the quench surface $t^2 - x^2  = T_0^2 \equiv \eta = 0$ as well as the post-quench conformal field theory. Further, the pre-quench state, which is assumed to be the vacuum state of the relativistic quantum field theory, is also preserved by Lorentz boosts. These properties imply that the post-quench state must also be invariant under $\xi$ translations. More precisely, multi-point correlation functions $\langle O_1 ( \eta_1,\xi_1) \cdots \rangle_{\text{CFT}}$ in the post-quench state can depend only on the differences $\xi_i - \xi_j$.

We now consider the one-point function of the stress tensor $\avg{ T_{\mu\nu} (\eta,\xi ) }_{\text{CFT}}$. (The energy density in particular is given by the component $T_{tt} (x,t)$ of the stress-energy tensor and will be recovered in what follows starting from the constraints imposed on the stress energy tensor in the conformal coordinate system.) This is independent of $\xi$ by the arguments above. Tracelessness and conservation then imply the general form
\begin{equation}
\begin{split}
\avg{ T_{\mu\nu} (\eta,\xi ) }_{\text{CFT}}  =  \begin{pmatrix} A & B \\ B & A \end{pmatrix} \;.
\end{split}
\end{equation}
where we note that the components above do not depend on $\eta$; this follows from the conservation operator identity $\nabla_\mu T^{\mu \nu} = 0$. This is a feature of 1+1 dimensions and does not hold in higher dimensions. Of course, $A$ and $B$ depend crucially on the details of the pre-quench state and the quench itself.

This may be further simplified in theories that are parity invariant, i.e. invariance under $x \to - x$ or equivalently $\xi \to - \xi$. This holds in many condensed matter systems and we assume it holds in the rest of the paper. As before, this symmetry is preserved by the quench and is therefore a symmetry of the post-quench state. Parity invariance together with translational invariance then implies that $\avg{ T_{\eta\xi} ( \eta , \xi ) }_{\text{CFT}} = B = 0$. With the form of the stress-energy tensor of the theory completely fixed up to a constant, we can revert back to the laboratory coordinates and read off the following result for the stress-energy tensor:
\begin{align}
\label{eq:Tres}
\avg{ T_{ab}(t,x) }_{\text{CFT}} = \frac{ f( m T_0)  }{ ( t^2 - x^2 )^2 } \begin{pmatrix} t^2+x^2 & -2xt  \\ -2xt  & t^2+x^2 \end{pmatrix}
\end{align}
where by dimensional analysis we have incorporated the constant as an appropriate scaling function above (recall that $m$ is the gap of QFT). This scaling function depends on the precise details of the QFT and CFT we are working with but there are some general claims that can be made regarding the limiting behavior of this function. First, in the limit of the mass $m \rightarrow 0$, the energy generated in the quench must go to zero since there is no perturbation in this limit. Thus, $f(mT_0) \rightarrow 0$ in the limit $m \rightarrow 0$ and keeping $T_0$ fixed. By corollary, this implies that for fixed $m$ and $T_0 \rightarrow 0$, that is as the quench approaches the luminal limit, the lab frame energy density $\avg{ T_{tt} }_{\text{CFT}} $ vanishes everywhere away from the light cone $t = \pm x$. 

Let us further note that the procedure works just as well for $T_0 \neq 0$. As is clear from Eq.~(\ref{eq:Tres}), the energy density at finite $x$ tends to zero as $\sim 1/t^2$ everywhere. This behavior can be attributed to conformal dilation of the energy density as follows. Excitations are created at all wave-vectors in $\xi$-coordinates, and these wave-vectors are preserved for subsequent time evolution in $\eta$. For $x \ll t$, we have $dx \sim t d \xi$---thus modes varying over a length scale $d x \sim T_0 d \xi$ at the initial time vary at a length scale $t d \xi$ at long times. This dilation of wave vectors causes the energy of massless modes to decrease as $\sim 1/t$. Moreover, the modes now occupy a larger volume, $\sim t$. These effects together result in a decrease in energy density as $\sim 1/t^2$. Finally, as a consequence, the energy density is $\sim \mathcal{O} \left[ 1/L^2 \right]$ in time $t \sim \mathcal{O} \left[ L \right]$, putting these quenches (with $T_0 \neq 0$) in the same class parametrically as the quench for $T_0 \rightarrow 0$ (although the latter is clearly faster), and faster than the uniform adiabatic algorithm which takes time $t \sim \mathcal{O} \left[ L^2 \right]$ to generate a state with exponentially-small energy density above the vacuum state. 

This completes our discussion of the general proof of the effectiveness of the cooling procedure for arbitrary CFTs. We next verify the arguments explicitly by showing that the quench in the Gaussian theory conform exactly to Eq. (\ref{eq:Tres}), and then discussing the protocol in the context of a model with non-linearities and infrared/ultraviolet cut-offs, the $O(N)$ model in the large-$N$ limit. 

\section{Quench in the Gaussian Theory}
\label{sec:Gaussianquench}

We now solve the quench in the case of a Gaussian scalar field and show that the energy density is explicitly of the form predicted by Eq.~(\ref{eq:Tres}). 

\subsection{Preliminaries}

The free massive scalar field is described by the action
\begin{equation}
\begin{split}
& S[\phi] =  - \frac{1}{2}  \int d^2 x \sqrt{-g}  \big[ g^{\mu\nu} \p_\mu \phi \p_\nu \phi + m^2 \phi^2  \big] \\
&~~ =  \frac{1}{2}  \int d\eta d\xi \big[  ( \p_\eta \phi )^2 -  ( \p_\xi \phi )^2 -  m^2 T^2_0 e^{2\eta}  \phi^2 \big].
\end{split}
\end{equation}
which leads to the equations of motion 
\begin{equation}
\big[ - \p_t^2 + \p_x^2 - m^2 \big] \phi = \big[ - \p_\eta^2 + \p_\xi^2 -  m^2 T^2_0 e^{2\eta}  \big] \phi  = 0. 
\label{eq:eom}
\end{equation}
We begin by working in the conformal coordinates.  The general solution to the equations of motion Eq.~\ref{eq:eom} is 
\begin{equation}
\begin{split}
\label{eq:udef}
\phi &= \int \frac{ dq }{ 2\pi } \left[ a_q  u_q  +  a^\dagger_q u_q^* \right]  ,  \\
u_{q}(\eta,\xi) &=  e^{i q \xi  } f_q(\eta) , \\
f_q(\eta) &= \frac{ \sqrt{\pi} }{ 2 } e^{ \frac{\pi}{2} q }   H^{(2)}_{i q } \left( m T_0 e^\eta  \right).
\end{split}
\end{equation}

where $H^{(2)}$ are Hankel functions of the second kind. The mode coefficients $a_q$, and $a^\dagger_q$ are elevated to operators in the quantum-mechanical setting, but we do not specify the commutation relations between these for the moment---these will be set by demanding that the field operators satisfy equal-time (real time $t$) commutation relations. The normalization of the modes satisfies
\begin{align}
\left( u_{q}    ,    u_{q'}  \right)_\eta &=  2\pi \delta(q - q')  , \qquad  \left( u_{q} , u^*_{q'}  \right)_\eta =  0 .
\end{align}
where the operation $(\cdot,\cdot)$ corresponds to the Klein-Gordon inner product defined by
\be
\left( \phi_1 , \phi_2 \right)_\eta =  i \int d\xi \left( \phi_1^*  \p_\eta \phi_2 - \phi_2  \p_\eta \phi_1^* \right) . 
\ee
 As a consequence of the above, the mode expansion coefficients can be extracted from a particular solution $\phi$ via the relations

\be
a_q = \left( u_q, \phi \right)_\eta, \; \; \; \; a^\dagger_q = \left(\phi, u_q \right)_\eta. 
\label{eq:brel}
\ee 

We could have also analyzed the problem in the laboratory frame where in the most general solution takes the form
\begin{equation}
\begin{split}
\label{eq:Udef}
\phi &= \int \frac{dk}{2\pi}  \left[ A_k  U_k  +  A^\dagger_k U^*_k \right], \\
U_{k} (t,x) &= e^{i k x } F_k(t), \\
F_k(t) &=  \frac{1}{\sqrt{2\omega_k}} e^{- i \omega_k t }, ~ \omega_k \equiv \sqrt{k^2 + m^2 } .
\end{split}
\end{equation}
As before the normalization is chosen so that
\begin{equation}
\begin{split}
\left( U_{k}    ,    U_{k'}  \right)_t &=  2\pi \delta ( k - k'  ), \\
\left( U_{k}  , U^*_{k'}  \right)_t &= 0, \\
\left( \phi_1 , \phi_2 \right)_t &=  i \int dx \left( \phi_1^*  \p_t \phi_2 - \phi_2  \p_t \phi_1^* \right), 
\end{split}
\end{equation}
and where the KG inner-product acts on a fixed time $t$-slice. Again, we may find the mode coefficients in this expansion via the relation 

\begin{align}
A_k = ( U_k , \phi)_t, \; \; \; \; A_k^\dagger = ( \phi , U_k )_t .
\end{align}

The two solutions for the field operator in Eqs.~(\ref{eq:udef}) and~(\ref{eq:Udef}) must obviously agree with each other and additionally satisfy the correct equal-$t$ commutation relations. 

In order that the solution of Eq.~(\ref{eq:Udef}) satisfy the equal $t$ commutation relations $ [ \phi (t, x), \partial_t \phi(t, x')  ] = i \delta (x - x')$, we require $ [ A_k, A^\dagger_{k'} ] = 2 \pi \delta (k - k')$. (In fact, the simplicity of the final commutation relations to ensure the correct equal-time commutation relations is a byproduct of the form of choosing the Klein-Gordon inner product for normalizing modes.) The usual Fock vacuum $\ket{\Omega}$ is represented by the vacuum of the bosons $A_k\,\forall\,k$. 

We now show that both the above conditions are satisfied for the solution in Eqs.~(\ref{eq:Udef}) for the choice 

\begin{align}
[a_q, a^\dagger_{q'}] = 2 \pi \delta ( q - q'), \; \; \; \;  a_q \ket{\Omega} = 0 \; \forall \,k
\end{align} 

This result is a happy accident, as was pointed out Fulling et al.~\cite{Fullingconformalvacuum} which occurs because translations in $\xi$ are an isometry of the space time. In fact, one can show that the solution even satisfies an equal-$\eta$ commutation relation $ [ \phi (\eta, \xi) , \p_\eta \phi (\eta, \xi')  ] = i \delta ( \xi - \xi' ) $.  The fact that the vacuum of bosons $A_k$ agrees with the vacuum of bosons, $a_q$ follows from the fact that the `positive-frequency' modes of the two sets of solutions can be expressed in terms of each other without the aid of using the `negative-frequency' or complex-conjugate solution. In particular, the following results hold 
\begin{equation}
\begin{split}
\label{Uurel}
u_{q}  &=  i \sqrt{2\pi} \int \frac{dk}{2\pi} \frac{1}{\sqrt{\omega_k}} e^{ i q \,\textrm{arcsinh} \frac{k}{m}  }  U_{k}, \\
U_{k} &= - i \sqrt{2\pi  } \int   \frac{dq}{2\pi} \frac{1}{\sqrt{\omega_k}} e^{-  i q \,\textrm{arcsinh} \frac{k}{m}  }   u_{q }.
\end{split}
\end{equation}
Using the identities 
\begin{equation}
\begin{split}
\int  \frac{dk}{2\pi} \frac{1}{ \omega_{k}  }  e^{ i ( q - q' ) \,\textrm{arcsinh} \frac{k}{m}} &= \delta ( q - q' ), \\
\int \frac{dq}{2\pi} e^{ i q \left[ \,\textrm{arcsinh} \frac{k}{m}  - \,\textrm{arcsinh} \frac{k'}{m} \right] }  &=  \omega_k \delta \left( k - k' \right), 
\label{eq:Aaidens}
\end{split}
\end{equation}
we can show
\begin{equation}
\begin{split}
A_{k  }  &= i \sqrt{2\pi}    \int  \frac{dq}{2\pi}  \frac{1}{\sqrt{\omega_k}} e^{ i q \,\textrm{arcsinh} \frac{k}{m} } a_{q}, \\ 
a_{q}  &= - i \sqrt{2\pi  } \int  \frac{dk}{2\pi} \frac{1}{\sqrt{\omega_k}}e^{-  i q\,\textrm{arcsinh} \frac{k}{m} } A_{k}.
\label{eq:Aarel}
\end{split}
\end{equation}
Thus, a state annihilated by all $A_k$ is concomitantly also annihilated by all $a_q$. 

\subsection{The Quench}

\begin{figure}[ht!]
\begin{center}
\includegraphics[width=3in]{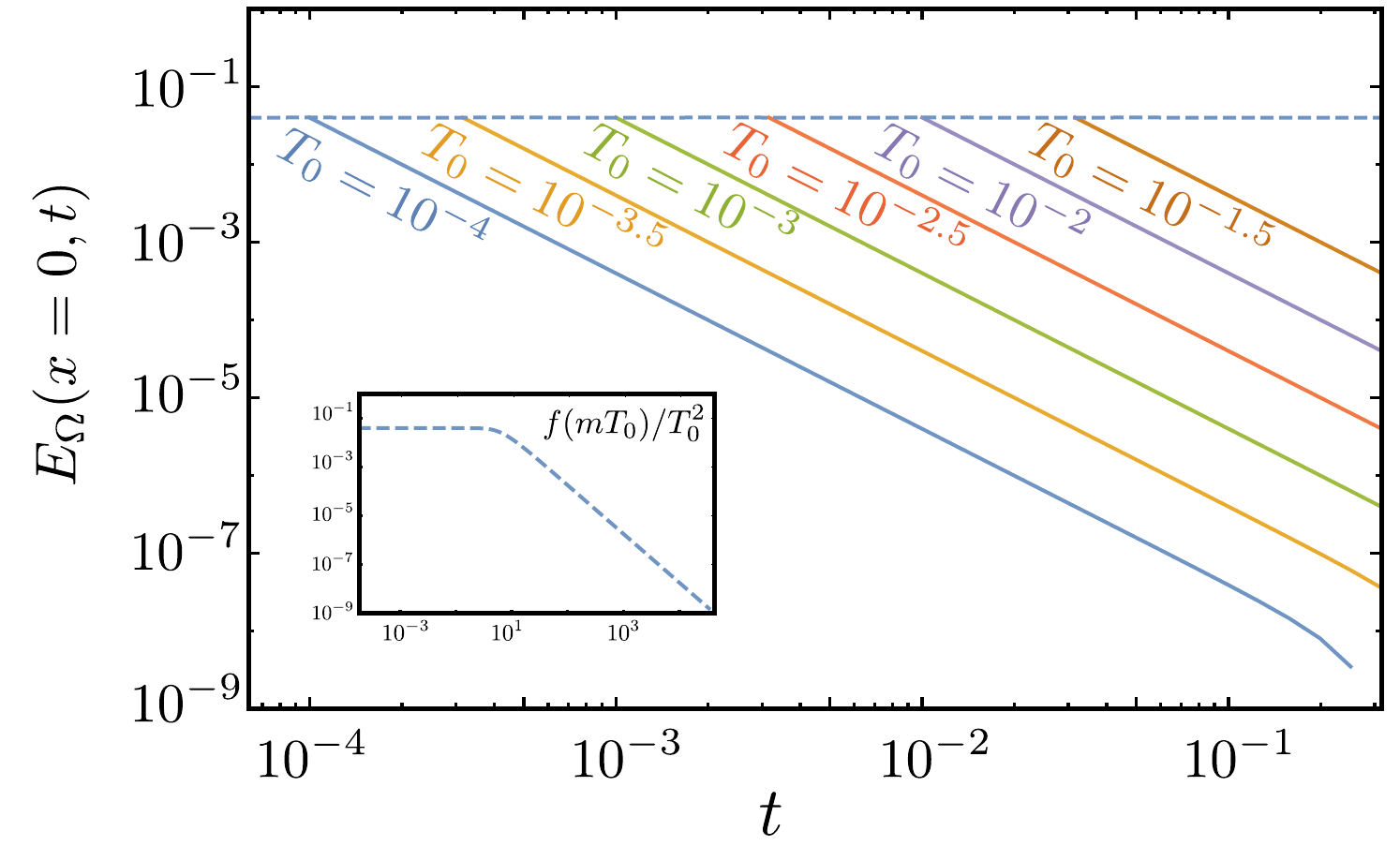}
\caption{The energy density at $x = 0$ is plotted as a function of time, $t$. For $T_0 \ll m^{-1} = 1$, the energy density at the origin at the precise instant of the quench is independent of $T_0$, and subsequently decays as $t^{-2}$. This agrees with expectations from Eq.~(\ref{eq:Tres}). In the inset, we examine the scaling function of Eq.~(\ref{eq:Tres}) by plotting $E_\Omega (x = 0, t = T_0)$. We see that $f(m T_0) \sim (m T_0)^2 $ for $T_0 \ll m^{-1}$, and $f(m T_0) \sim m T_0$ for $T_0 \gg m^{-1}$. Note that the former implies that $\epsilon = 2$ for the Gaussian theory.}
\label{fig:Gaussian1}
\end{center}
\end{figure}

\begin{figure*}
\begin{center}
\includegraphics[width=6in]{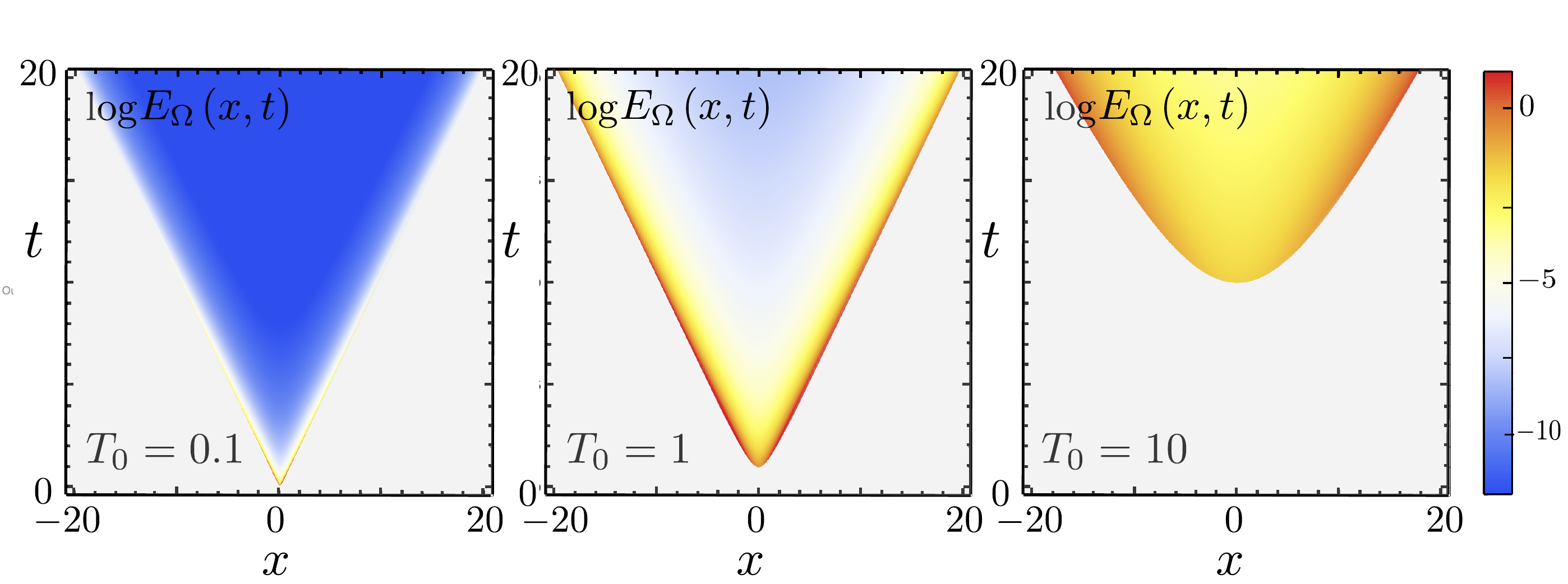}
\caption{The energy density $E_\Omega(x,t)$ of the Luttinger liquid (or Gaussian scalar field) with initial mass $m = 1$, Luttinger parameter $K = 1$, is plotted after a quench from a state with a finite mass for different quench trajectories parameterized by $T_0$. The quench is most effective in the limit $T_0 \rightarrow 0$, but results in cooling at long times for all $T_0$.}
\label{fig:Gaussian2}
\end{center}
\end{figure*}

We now calculate the properties of the system after the quench. Recall that the initial mass of the system is $m$ and at $\eta = 0$, it is quenched to ${\tilde m} \rightarrow 0$ (all post-quench quantities are capped by a tilde).  We then have the equations
\begin{equation}
\begin{split}
( \Box - m^2 ) \phi = 0~(\eta < 0), \\
( \Box - {\tilde m}^2 ) {\tilde \phi} = 0~(\eta > 0), \\ 
\{ \phi  - {\tilde \phi} , \p_\eta \phi - \p_\eta {\tilde \phi}  \}   |_{\eta = 0} = \{ 0 , 0 \} .
\label{eq:cont}
\end{split}
\end{equation}
[Here $\Box \equiv e^{- 2 \eta} ( \p^2_\xi - \p^2_\eta )$.] The last equation of the above corresponds to the two continuity of the amplitude and time-derivative of the field operators. The first two equations are solved by
\begin{equation}
\begin{split}
\phi &= \int \frac{dq}{2\pi} \big[ a_q u_q + a_q^\dagger u_q^* \big], \\
{\tilde \phi} &= \int  \frac{dq}{2\pi} \big[ \tilde a_q \tilde u_q + \tilde a_q^\dagger \tilde u_q^* \big], 
\end{split}
\end{equation}
where the modes $\tilde{a}_q$ are defined as $a_q$ but with the mass set to $\tilde{m}$. The last two equations can be used to relate the pre- and post-quench mode coefficients. To do this, we evaluate Eq.~(\ref{eq:brel}) at $\eta=0$ where using the last equation of Eq.~(\ref{eq:cont}), we can replace $\phi \leftrightarrow {\tilde \phi}$. Thus,
\begin{equation}
\begin{split}
a_q  &=  ( u_q , \tilde{\phi}  )_{\eta = 0} =  \alpha_q {\tilde a}_q + \beta_q {\tilde a}^\dagger_{-q}, \\ 
\tilde{a}_q &=  ( {\tilde u}_q  , \phi  )_{\eta_0} = \alpha^*_q a_q - \beta_q  a^\dagger_{-q}  .
\end{split}
\end{equation}
with Bogoliubov coefficients
\begin{equation}
\begin{split}
\alpha_q &= i  [ f^*_q \tilde{f}'_q - \tilde{f}_q f'^*_q  ]_{\eta = 0} , \\
\beta_q &= i  [ f^*_q \tilde{f}'^*_q - \tilde{f}^*_q f'^*_q ]_{\eta =0}. 
\end{split}
\end{equation}

where the functions $f_q$ were defined in Eq.~(\ref{eq:udef}). Note that the commutation relations of the bosons $a_q$ and $\tilde{a}_q$ require $\abs{\alpha_q}^2 - \abs{\beta_q}^2 = 1$. As a further check, one can show easily that for $\tilde{m} = m$, $\alpha_q = 1$, and $\beta_q = 0$. We may thus represent the post-quench field operator result
\begin{equation}
\begin{split}
\tilde{\phi} &= \int \frac{dq}{2 \pi} \left[ a_q \tilde{\gamma}_q + a^\dagger_q \tilde{\gamma}^*_q \right],  \\
\tilde{\gamma_q} &= e^{i q \xi} \left[ \alpha^*_q \tilde{f}_q - \beta^*_q f^*_q \right] . 
\end{split}
\end{equation}
The coordinate-invariant representation of the stress-energy tensor of the Gaussian scalar field theory in $d=2$ is
\begin{align}
T_{\mu \nu} = \nabla_\mu \tilde{\phi} \nabla_\nu \tilde{\phi} + \frac{1}{2} g_{\mu \nu} [-\nabla^\gamma \tilde{\phi} \nabla_\gamma \tilde{\phi} + \tilde{m}^2 \tilde{\phi}^2 ] . 
\end{align}
The expectation value of this tensor above the vacuum is defined by $\avg{T_{\mu \nu}} = \matrixel{\Omega}{\!\!\vs{:} \! T_{\mu \nu}\!\vs{:} \!\!}{\Omega}$, where we have introduced normal ordering such that the expectation value of the stress-energy tensor in vacuum is identically zero. Using the explicit formulae of the previous section, we find
\begin{equation}
\begin{split}
\avg{T_{\eta\eta}} &= \avg{ T_{\xi\xi} } = \int \frac{dq}{4 \pi} \bigg[ (\abs{\p_\eta \tilde{\gamma}_q}^2 - \abs{\p_\eta \tilde{u}_q}^2) \\
&\qquad \qquad \qquad \qquad + (\abs{\p_\xi \tilde{\gamma}_q}^2 - \abs{\p_\xi \tilde{\gamma} }^2) \\
&\qquad \qquad \qquad \qquad  + \tilde{m}^2 (\abs{\tilde{\gamma}_q}^2 - \abs{\tilde{u}_q}^2) \bigg],  \\
\avg{ T_{\eta\xi} } &= \int \frac{dq}{2 \pi} \bigg[ \p_\mu \tilde{\gamma}_q \p_\nu \tilde{\gamma}^*_q - \p_\mu \tilde{u}_q \p_\nu \tilde{u}^*_q \bigg]
\label{eq:Tconformal}
\end{split}
\end{equation}
Using the explicit formulae derived in this section, we find $\avg{ T_{\eta\xi} } = 0$ since the corresponding integrand is odd in $q$. With the value of $\avg{ T_{\mu \nu} } $ determined, we may perform a coordinate transformation to easily show that $\avg{ T_{ab} } $ that satisfies Eq.~(\ref{eq:Tres}), proving the assertion for the Gaussian theory. While analytic formulas are available, they are unwieldy for direct evaluation. In Fig.~\ref{fig:Gaussian1}, we numerically evaluate the scaling function $f(m T_0)$ of Eq.~(\ref{eq:Tres}). In Fig.~\ref{fig:Gaussian2}, we show the cooling effect graphically with a heat map in space-time. 


\subsection{Quench from excited states\label{sec:excited}}

We now show that at least for the Gaussian theory, the quench protocol works even when starting with excited states, in the sense that it maps an excited state of the massive theory to one with the same population in the massless theory, with no entropy production in the bulk. 

In particular, let us assume that we start in a state described by a large gap $m$ such that the dispersion of relevant modes may be neglected. (For instance, if we smooth the protocol everywhere on a time-scale $\tau \sim m^{-1}$, the quench is a non-adiabatic process only for modes with energy $\omega_k \lesssim \tau^{-1}$, or momenta $k \lesssim m$ for which we may ignore the dispersion.) In this case, the mode occupation of the laboratory-coordinate plane-wave modes is $\avg{A_k A^\dagger_{k'}} = 2 \pi \delta_{k,k'} n_k\approx 2 \pi n_B (m / k_B T) \; \forall \; k \ll m$. Here $n_B$ is the usual thermal boson population function. Now, using the results of Eqs.~(\ref{eq:Aaidens}) and~(\ref{eq:Aarel}), this initial state can be equivalently represented in terms of a uniform population of the conformal modes $a_q$, satisfying $\avg{a^\dagger_q a_{q'}} = 2 \pi \delta_{q,q'} n_B (m / k_B T)$. 

\begin{figure}
\centering
\includegraphics[width=0.8\columnwidth]{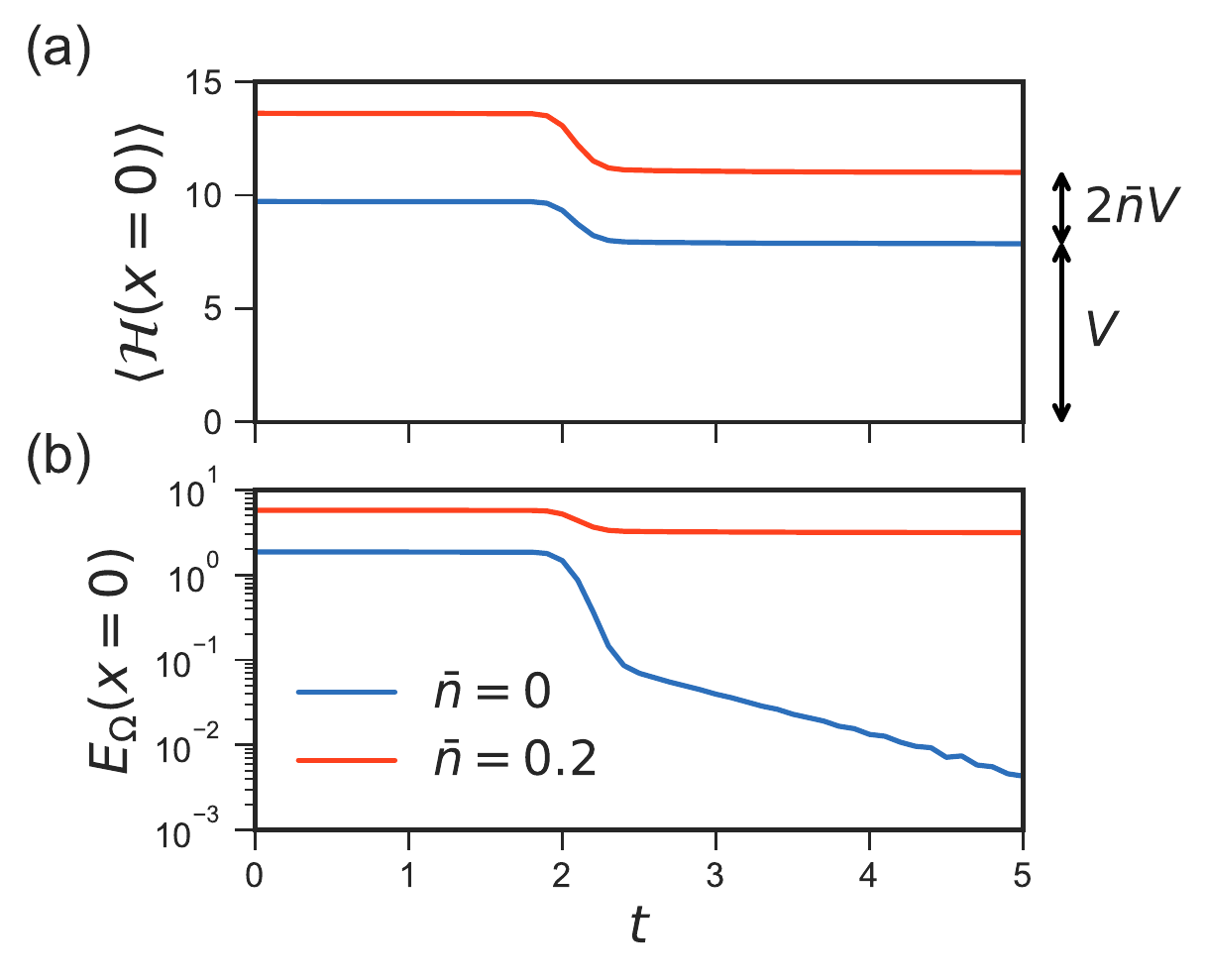}
\caption{
Illustration of the protocol starting from the vacuum ($\bar{n}=0$) and from a uniformly populated excited state ($\bar{n} = 0.2$) in the Gaussian theory. 
(a) Energy density at $x=0$ as a function of $t$, showing saturation to $V$ and $(1+2\bar{n})V$ in the two cases.
(b) Energy density {above the final vacuum}, $E_\Omega$, in logarithmic scale.
\label{fig:excited}}
\end{figure}

We would now like to find the mode population of the post-quench massless modes $\tilde{a}_q$, and more importantly, the population of the massless plane-wave modes $\tilde{A}_q$. (The modes $\tilde{A}_q$ are defined as the modes $A_q$ but with a mass $\tilde{m} \rightarrow 0$). The result is surprisingly simple: the population of modes $\tilde{A}_q$, is precisely given by the population of the initial massive plane-wave modes $A_q$, that is, $\avg{\tilde{A}^\dagger_q \tilde{A}_{q'}} = \avg{A^\dagger_q A_{q'}} = 2 \pi \delta_{q,q'} n_B (m / k_B T)$. To see this, first note that, as a corollary, the post-quench conformal mode population is also $\avg{\tilde{a}^\dagger_k \tilde{a}_{k'}} = 2 \pi \delta_{k,k'} n_B (m / k_B T)$. Next, if we denote the pre-quench state by $\ket{E}$ and the post-quench state with the population $\avg{\tilde{a}^\dagger_k \tilde{a}_{k'}}$ as noted above by $|\tilde{E}\rangle$, it follows that 

\begin{align}
 \matrixel{E}{T_{\mu \nu}}{E} &=  \langle \tilde{E} | T_{\mu \nu}| \tilde{E}\rangle \nonumber \\
& + \left[2 n_B \left(\frac{m}{k_B T}\right) + 1\right] \matrixel{\Omega}{\!\! \vs{:} \! T_{\mu \nu} \! \vs{:} \!\!}{\Omega}\;.
\label{eq:Texcited}
\end{align} 
The coordinate transformation of this result then reveals, as per previous analysis, that the energy density of the post-quench system now relaxes to the energy density of the state $\tilde{E}$, which is categorized by the exact same population of modes $\sim n_B (m / k_B T)$ as the massive modes before the quench. (See Fig.~\ref{fig:excited} for a numerical verification of this result in the Gaussian case.)

This result is particularly significant from an experimental point of view as it is relatively simpler to prepare a state with a temperature $T \ll m/k_B$ such that the initial mode population is exponentially suppressed. Further, note that the same population of modes before and after the quench naively implies that there is precisely no entropy creation in this rather violent quench process---however, we must recall that there are singular features along the lines $x = \pm t$ prevalent in the second term of Eq.~(\ref{eq:Texcited}). 
Nonetheless, given the spatial locality of these singular features, we expect all the entropy production to be channeled along these singular lines, much like all the energy production was seen to be concentrated on these lines for the vacuum quench. We leave a more detailed exploration of the entropy production in the Gaussian theory, as well as a more general treatment in the interacting setting, for future work.

\section{Numerical simulations}
\label{sec:numerics}

In the following we present numerical simulations of the quench in an interacting quantum theory, and also probe the effect of various non-idealities that result in the loss of conformal invariance in realistic systems. We study these questions in the context of the quantum $O(N)$ model in the large-$N$ limit which admits a simple mean-field description and allows for extensive numerical simulation. While this is not a conformal field theory, the gap can in principle be made arbitrarily small---when this gap is made smaller than other cutoffs, the system is approximately critical, and can be used to test our theoretical analysis in an interacting context. 

\subsection{The $O(N)$ model in the large-$N$ limit}

In $d=1$, the Hamiltonian of the $O(N)$ model reads
\be
H = \frac{1}{2} \int d x \left( \abs{\vs{\Pi}}^2 + \abs{\p_x \vs{\Phi}}^2 + r \abs{\vs{\Phi}}^2 + \frac{\lambda}{2 N} \abs{\vs{\Phi}}^4 \right)
\ee
where $\vs{\Phi}$ and $\vs{\Pi}$ are canonically conjugate $N$-component fields, 
satisfying 
$$[ \Phi_i (t, x), \Pi_j (t, x') ] = i \delta ( x - x') \delta_{ij},$$
$\abs{\vs{\Phi}}^2 = \sum_i \Phi_i^2$, and $\abs{\vs{\Phi}}^4 = {(\abs{\vs{\Phi}}^2)}^2$.
In the limit $N \rightarrow \infty$, the Hamiltonian is amenable to a mean-field treatment which reduces the problem to that of a self-consistent Gaussian scalar field theory. Within the mean-field ansatz which is exact in the limit $N \rightarrow \infty$ (and following Ref.~\cite{chandrancoarsening,Smacchia2015dQPTinlargeNON}), we note
\begin{align}
H_{\text{eff}} (t) &= \frac{1}{2} \int dx\ \left[ \abs{\vs{\Pi}}^2 + \abs{\partial_x \vs{\Phi}}^2 + m_{\text{eff}}^2 \abs{\vs{\Phi}}^2 \right], \nonumber \\
m^2_{\text{eff}} (t, x) &\equiv r + \lambda \left\langle\frac{\abs{\vs{\Phi} (t, x)}^2}{N}\right\rangle. 
\end{align}
which is a Gaussian theory with a self-consistently renormalized mass term. Since all components of $\vs{\Phi}$ are equivalent and independent in this limit, we need only work with a single one of them, $\phi(t,x)$, replacing $\avg{\abs{\vs{\Phi} (t,x) }^2}/N$ by the expectation value $\avg{\phi^2 (t,x)}$. The resulting equations of motion are non-linear but efficiently solvable on a classical computer.

In what follows, we work solely in laboratory frame coordinates $(t,x)$. We regularize the equations of motion by imposing infrared and ultraviolet cutoffs. The former is achieved by putting the system on a finite segment of length $L$ with periodic boundary conditions, $\phi(t,x+L) \equiv \phi(t,x)$. The latter comes in the form of a momentum cutoff, $|k| < \Lambda/2$, with $\phi_{k+\Lambda} (t) \equiv \phi_k(t)$. In practice this reduces the continuum field theory to a 1D lattice of $M \equiv L\Lambda/2\pi$ sites with periodic boundary conditions. Lattice sites correspond to momenta $k_n = \frac{2\pi}{L} n$, with $n \in \{-M/2+1, \dots,  M/2 \}$.

We perform a mode expansion of the field $\phi$:
\begin{equation}
\phi(t,x) = \sum_p f_p(t,x) a_p + f_p^\ast (t,x) a_p^\dagger \;,
\label{eq:pre_modeexpansion}
\end{equation}
where the $\{a_p\}$ are a set of annihilation operators for the initial vacuum labelled by momentum $p$, 
and the coefficients $f_p$ are called mode functions. 
In the presence of spatial translation invariance, these depend on position $x$ as $f_p(x) \sim e^{-ipx}$.
This does not hold when translation invariance is broken, as in our spatiotemporal quench protocol.
The Fourier-transformed field $\phi_k(t) \equiv M^{-1/2} \sum_x \phi(t,x) e^{ikx}$ can be written as
\begin{equation}
\phi_k(t) = \sum_p f_{p,k}(t)  a_p + f^\ast_{p,-k}(t) a_p^\dagger \;.
\label{eq:modeexpansion}
\end{equation}
where $f_{p,k}(t) \equiv M^{-1/2} \sum_x f_p(t,x) e^{ikx}$.
We call $f_{p,k}$ the {\it mode function matrix}. It is an $M\times M$ matrix; both its indices are momenta.
In the presence of translation invariance (see e.g. the treatment in Ref.~\cite{chandrancoarsening}) the matrix is diagonal, and the dynamics is fully described by the mode vector $f_k$.
In the present case, $f_{p,k}(0)$ is initially diagonal, but off-diagonal entries are populated dynamically over the course of the quench. 

Upon decomposing $\phi(t,x)$ as in Eq.~\eqref{eq:modeexpansion}, 
the equation of motion 
\begin{equation}
(\partial_t^2 -\partial_x^2 + r(t,x) + \lambda \langle \phi^2(t,x) \rangle ) \phi(t,x) = 0 \;
\label{eq:ON_eom}
\end{equation}
maps to a system of $M^2$ non-linear, time-dependent, second-order ODEs in the mode functions $f_{p,k}(t)$. Canonical commutation relations $[\phi_p, \phi_q] = [\pi_p, \pi_q] = 0$, $[\phi_p, \pi_q] = i \delta_{p,q}$ are mapped to the matrix equations
\begin{equation}
\Imag(f^\dagger f) = \Imag(g^\dagger g)  = 0,
\quad 
\Imag(f^\dagger g) = - \frac{1}{2} I \;,
\end{equation}
where $I$ is the $M\times M$ identity matrix and $g = \dot{f}$.
The initial conditions are given by
$$
f_{p,k}(0)= \sqrt{\frac{1}{2\Omega_k}}\delta_{p,k},
\qquad
g_{p,k}(0) = -i \sqrt{\frac{\Omega_k}{2}} \delta_{p,k}\;,
$$
with $\Omega_k  = \sqrt{k^2 + m^2_{\rm eff}}$. 
The effective mass $m_{\rm eff}$ is determined self-consistently by demanding that $m^2_{\rm eff} = r+\lambda \avg{\phi^2}$:
\begin{equation}
m^2_{\rm eff} = r + \frac{\lambda}{2M} \sum_{n=-M/2}^{M/2-1} \frac{1}{\sqrt{k_n^2 + m^2_{\rm eff}}} \;,
\end{equation}
with $k_n = 2\pi n /L$.
The effective mass is always positive, but approaches 0 as $r\to-\infty$. 

We implement the quench protocol by varying the parameter $r$ in a space-time dependent with the form
\begin{equation}
r(t,x) = r_0 + (r_1-r_0) \sigma\left(t-\sqrt{x^2+T_0^2}\right) \;,
\label{eq:rxt}
\end{equation}
where $\sigma(t) = \frac{1}{2} (1+\tanh(t/\tau))$ is a step function smoothed over a time scale $\tau$. 
The limit $\tau \to 0$ yields the instantaneous quench discussed analytically in Sec.~\ref{sec:gened} and \ref{sec:Gaussianquench}. 

\begin{figure}
\centering
\includegraphics[width=\columnwidth]{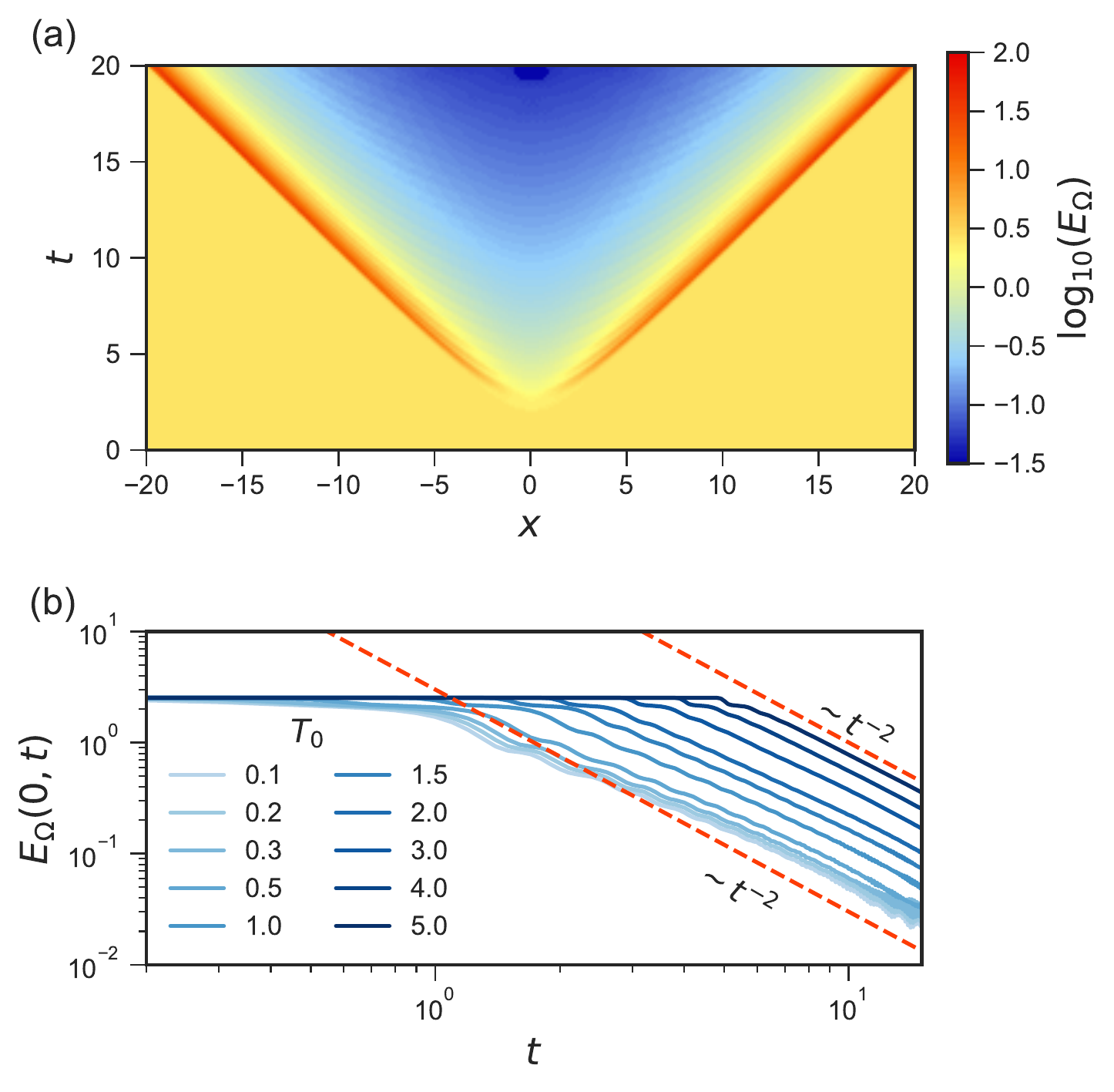}
\caption{
Simulated quench in the large-$N$ $O(N)$ model.
(a) Color plot of the energy density in spacetime for $T_0 = 2$.
(b) Energy density above the final vacuum $E_{\Omega}$ at $x=0$ exhibits $1/t^2$ relaxation across a range of values of the quench parameter $T_0$.
The squared-mass parameter $r$ is quenched from $r_0 = 1$ to $r_1 = -10$.
The interaction, with strength $\lambda = 10$, renormalizes the initial mass squared to ${m}_{\rm eff}^2 \simeq 1.6$ and the final one to ${m}_{\rm eff}^2 \simeq 1.6\times10^{-6}$, giving a nearly massless theory.
The IR cutoff (system size) is $L = 40$, the UV cutoff is $\Lambda = 20 \pi$. 
The simulation includes $M = L\Lambda/2\pi = 400$ modes.
\label{fig:ON}}
\end{figure}

We numerically integrate the equations of motion [Eq.~(\ref{eq:ON_eom}), with the form of $r(x,t)$ as in Eq.~(\ref{eq:rxt})] using a fourth-order Runge-Kutta method. The evolved mode function matrix $f_{p,k}(t)$ is used to calculate the energy density above the vacuum of the final theory, $E_{\Omega}(x,t)$ at various time steps and compared to the analytical predictions. An example is shown in Fig.~\ref{fig:ON}(a), in the form of a heat map in space-time. We clearly see the hot region near the quench front $t^2-x^2 = T_0^2$, as well as the cooling effect away from the front.
To better understand the nature of this cooling, in Fig.~\ref{fig:ON}(b) we show the energy density at the center of the system, $E_\Omega(x=0, t)$, as a function of time.
Eq.~\eqref{eq:Tres} predicts this should fall off as $t^{-2}$ for an instantaneous quench to a CFT. 
Despite the presence of a variety of finite cutoffs that break conformal invariance (the small but finite mass, the IR and UV cutoffs $L$, $\Lambda$, and the smoothing time $\tau$), we see good agreement with the predicted scaling.
This confirms the validity of the argument presented in Sec.~\ref{sec:gened} away from the non-interacting limit considered in Sec.~\ref{sec:Gaussianquench}, and also indicates its robustness to small violations of the assumptions.
The latter aspect is important for the experimental applicability of the protocol, and we investigate it more thoroughly in the following.


\subsection{Effect of finite cutoffs}

In a realistic implementation of the protocol, the massless modes in the final theory would likely have a phonon-like dispersion: linear near $k=0$, but highly non-linear, or even flat, sufficiently far from $k = 0$.
As a simple model for this, we consider the dispersion
\begin{equation}
\Omega_k^2 = m_{\rm eff}^2 + \left( \frac{\Lambda}{\pi} \right)^2 \sin^2 \left( \frac{\pi}{\Lambda} k \right) \;.
\label{eq:phonon_dispersion}
\end{equation}
Here $\Lambda$ is the UV cutoff, defined by $|k| < \Lambda/2$.
It can be though of as arising from a microscopic lattice spacing $a$ via $\Lambda = 2\pi/a$, giving $|k|<\pi/a$.
If $m_{\rm eff}$ is quenched from a positive value to zero, after the quench one gets modes that propagate at the speed of light only for $|k| \ll \Lambda$.
For these modes, the cooling argument is expected to hold, with the energy production being confined in spacetime near the quench front.
But for modes with $|k| \sim \Lambda$, that have group velocities much smaller than 1, the argument is expected to fail.
This seemingly implies that the cooling protocol is inapplicable in systems with phonon-like dispersions of the type in Eq.~\eqref{eq:phonon_dispersion}.

\begin{figure}
\centering
\includegraphics[width=\columnwidth]{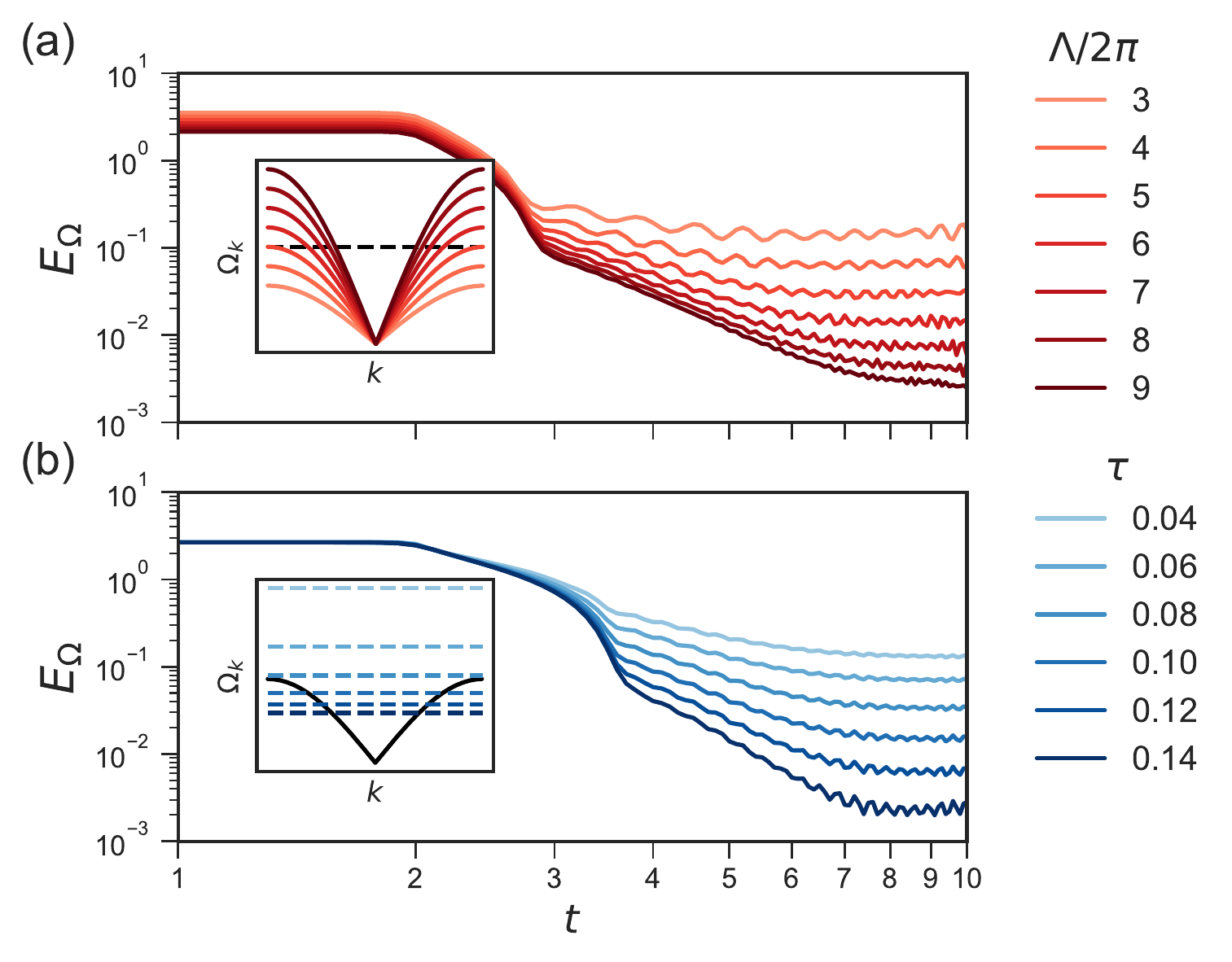}
\caption{Effect of cutoffs on quench in the free theory with phonon-like dispersion, Eq.~\eqref{eq:phonon_dispersion}. 
$r$ is quenched from $r_0 = 10^2$ to $r_1 = 10^{-2}$.
The IR cutoff is $L = 40$. Interactions are absent ($\lambda = 0$) and $T_0 = 2$.
(a) Effect of increasing UV cutoff $\Lambda$ at fixed $\tau = 0.1$. 
The state relaxes down to a finite asymptotic energy density that decreases with $\Lambda$. 
Inset: the dispersion $\Omega_k$ vs $k$ for the values of $\Lambda$ considered (solid lines), along with the value of $\tau^{-1}$ (dashed line).
(b) Effect of increasing the smoothing time $\tau$ at fixed $\Lambda = 12\pi$. 
Inset: the dispersion $\Omega_k$ vs $k$ (solid line), along with the values of $\tau^{-1}$ considered (dashed lines).
\label{fig:phonon}}
\end{figure}

The key to overcoming this issue lies in the possibility of tuning the smoothing time, $\tau$, of the quench.
Indeed, a mode with frequency $\Omega_k \gg \tau^{-1}$ will see the quench as an {\it adiabatic} process, and thus will remain in its vacuum state.
Only modes with low enough frequency will be affected by a sudden quench, and by tuning $\tau$ one can ensure those modes are within the linear dispersion regime, $\Omega_k \simeq |k|$.
This is illustrated in Fig.~\ref{fig:phonon}, where we present results of numerical simulations in the free theory ($\lambda = 0$) with the dispersion \eqref{eq:phonon_dispersion} with varying values of the UV cutoff $\Lambda$ and smoothing time $\tau$.

In particular, Fig.~\ref{fig:phonon}(a) shows a sweep over $\Lambda$ at fixed $\tau$.
The energy density at $x=0$ relaxes down to a finite asymptotic value above the vacuum, due to the excitations carried by large-$k$, sub-luminal modes; this value approaches 0 as $\Lambda$ is increased and sub-luminal modes are pushed past the $\tau^{-1}$ cutoff, crossing over into an adiabatic regime.
Fig.~\ref{fig:phonon}(b) shows instead a sweep over $\tau$ at fixed $\Lambda$,
confirming the same picture.
This latter scenario is also the one most relevant to experiment.
The UV cutoff $\Lambda$ is unlikely to be controllable in many implementations of 1D quantum systems. 
On the contrary, we can expect any experimental platform capable of implementing the quench protocol here described to have a reasonable degree of control over the smoothing time $\tau$.


\section{Higher dimensions}
\label{sec:higherd}

In this section, we consider a generalization of the above argument to $d>1$. In particular, we consider a hyperbolic quench protocol in the directions $(t,x)$, independent of the remaining spatial directions $(x_2,\cdots,x_d)$---such a protocol is chosen over a more symmetric radial quench because it is advantageous to work with the same special conformal coordinates as in the $d = 1$ case. We show that the stress-energy tensor is not completely constrained by the argument of symmetry and conformal invariance as in the $d = 1$ case. This does not of course rule out the possibility that the quench is effective in preparing low-energy states even in higher-dimensions. For instance, the quench protocol works in the Gaussian case in higher dimensions; we discuss this later. 

We denote the laboratory coordinates by $x^a = (t,x,x_2, \cdots , x_d )$ and the conformal coordinates denoted by $x^\mu = (\eta, \xi, x_2 , \dots , x_d)$. Then, assuming parity invariance, isotropy along the transverse directions, invariance under $\xi$-translations, conservation and tracelessness, the stress tensor takes the form
\begin{equation}
\begin{split}
\avg{ T_{\mu\nu} } &=\text{diag} \bigg(  A(\eta)  , A (\eta) - A'(\eta)  ,  \\ 
&\qquad \qquad \quad   \frac{e^{-2\eta} A'(\eta)}{d-1}  , \dots , \frac{  e^{-2\eta}  A'(\eta)   }{d-1}\bigg)  . 
\end{split}
\end{equation}
Thus, we are left with some particular function $A(\eta)$ whose $\eta$-dependence cannot be constrained by symmetry and conservation laws. For the energy density in the laboratory coordinates, we find the form
\begin{equation}
\begin{split}
\avg{ T_{tt} }  &= \frac{m^{d-1} }{(t^2 - x^2)^2 }  \bigg[ f(\eta , m T_0)  ( t^2 + x^2 )  \\
&\qquad \qquad \qquad \qquad \quad  -  \p_\eta f(\eta , m T_0)  x^2 \bigg], \\ 
\eta &= \frac{1}{2} \text{log} \bigg[ \frac{T^2_0}{t^2 - x^2} \bigg].
\end{split}
\end{equation}
where we have added factors of $m$ by dimensional analysis expressing $A(\eta)$ in terms of a dimensionless scaling function $f(\eta, m T_0)$. 

First note that due to the undetermined $\eta$-dependence, we cannot make any general claims about cooling for arbitrary $T_0$. We can examine the limit $T_0 \rightarrow 0$ by considering the limit $m \rightarrow 0$ instead. In this limit, we must obtain $\avg{T_{tt} } \to 0$. For $d > 1$, this is achieved as long as $f(\eta, mT_0)$ stays finite for $m\to 0$, or diverges too slowly to compensate the $m^{d-1}$ mass dependence. Thus, we cannot guarantee that the energy density relaxes to zero even in the limit $T_0 \rightarrow 0$.  

As mentioned above, this does not rule out the possibility that the quench does in fact work in higher dimensions. Here we show that the quench is in fact effective in cooling for the Gaussian theory. The $(d+1)$-dimensional free theory in this protocol is equivalent to a set of decoupled wires parametrized by $k_\perp$ undergoing the original $(1+1)$-dimensional protocol with all but a set of zero measure having $k_\perp^2>0$, thus giving rise to a massive-to-massive quench. Note again that the quench generates excitations with all wave-vectors in the conformal spatial coordinate $\xi$. In the free field theory, the wave-vectors of these modes are preserved over subsequent time-evolution. Following the same argument described in the second last paragraph of section \ref{sec:gened}, we find that the dilation of wave vectors at large $t$ causes the energy of massive modes to approach the finite value of the final ($k_\perp$-dependent) mass (This is contrast to massless modes whose energy approaches zero as $\sim 1/t$). Together with an increase in the occupation volume of these modes as $\Delta x \sim t$, we find a decrease in energy density as $\sim 1/t$. Thus, the quench results in cooling to a vacuum state in higher dimensions, as well.

\section{Summary and Discussion}
\label{sec:conclusions}

In this work, we studied a new kind of spatio-temporal quench protocol that can be utilized for the purpose of rapidly preparing ground states of critical models in one dimension. In particular, we studied the quench from the ground state of a gapped relativistic quantum system, closing the gap along the space-like trajectory $t^2 - x^2 = T^2_0$. We showed that such a quench causes the system to relax to the final vacuum everywhere except on the singular lines $x = \pm t$. This work extends our vocabulary of solvable spatio-temporal quenches in one-dimensional systems going beyond the previously studied case of single-velocity superluminal quenches. More importantly, it provides a robust \emph{geometric} argument for the validity of the findings in Ref.~\cite{AgarwalGroundStatePrep} to the case of general \emph{interacting} critical systems. These observations make the protocol we present particularly useful for preparing low-energy states in one-dimensional systems that may be described by a low-energy Luttinger liquid theory, or more generally by strongly-coupled conformal field theories. We confirmed these predictions by concretely studying the quench for the Gaussian scalar field theory, and numerically in an interacting setting by simulating the quench in the quantum $O(N)$ model in the large-$N$ limit. 

In this work we also considered quenches in the scalar free field theory starting from an excited state, finding that the mode populations of massive bosons (which in an experimental setting would be suppressed exponentially for a temperature smaller than the mass) directly translates to the mode population of the post-quench massless bosons. Thus, the quench process also appears to conserve entropy (again barring the singular lines $x = \pm t$). We leave a more careful treatment of this result for future work, where we also aim to explore such quenches from excited initial states in an interacting setting. 

Finally, we showed that the above arguments do not straightforwardly extend to the higher dimensional case where we find the stress-energy tensor cannot by constrained by symmetry arguments alone. This is perhaps natural given the uniqueness of one-dimensional systems where scattering is heavily constrained by the limited phase space available, and thus operates in a very different way to higher dimensional systems. 
However, we know that the quench protocol can be used in higher dimensions for the purposes of cooling (based on previous work in Ref.~\cite{AgarwalGroundStatePrep}) by introducing limited adiabaticity in the way of a finite time-scale $\tau$ over which the superluminal quench is smoothed. A numerical investigation of the efficacy of this method in creating the ground state of the Hubbard model at half-filling, which is of immense experimental importance especially for related studies in ultracold atoms, is also left for future work. We also mention in passing that progress may be made in understanding such quenches in higher dimensions by exploring whether insights from the Unruh result~\cite{Unruhoriginal,unruheffectandapplications} (for correlations of the system on hyperbolic time-like hypersurfaces) can be used in our computations which involve propagation of the system in time following a quench on related space-like hypersurfaces.    

Finally, from an experimental point of view, our findings may be investigated in systems of ultra-cold atoms trapped in flat-band potentials~\cite{HadzibabicUniformBose,mukherjee2015fermi}, besides arrays of Josephson Junctions~\cite{HavilandArray,UstinovJJArray}, and ion traps~\cite{blatt2012quantumsimulation,MBLTrappedIonMonroe,rajabi2018dynamic}. Experiments on atom chips~\cite{Gring,langen2015ultracold} studying in particular the dephasing between halves of a split quasi-one-dimensional condensate have been used to investigate spatially uniform, instantaneous quenches of the mass in a Luttinger liquid setting. A spatio-temporal quench of the sort investigated here may be created in these systems by splitting the quasi-condensate along the space-time trajectories discussed, and appear to be a promising experimental candidate for realizing the physics discussed here. 

\section{Acknowledgements}

The authors would like to thank Eugene Demler, Andrew Green, Juan Maldacena, Ivar Martin, Emanuele Dalla Torre, Frank Pollmann, and Ruben Verresen for insightful comments on this work. PM gratefully acknowledges support from DOE grant DE-SC0009988. This research was also funded by DOE-BES Grant No. DE-SC0002140 (MI, KA and RNB), and UK Foundation (KA). KA and RNB acknowledge the hospitality of KITP, UCSB during their program on Dynamics of Quantum Information during which this manuscript was finalized. SLS acknowledges support from US Department of Energy grant No. DE-SC0016244.


\begin{thebibliography}{62}%
\makeatletter
\providecommand \@ifxundefined [1]{%
 \@ifx{#1\undefined}
}%
\providecommand \@ifnum [1]{%
 \ifnum #1\expandafter \@firstoftwo
 \else \expandafter \@secondoftwo
 \fi
}%
\providecommand \@ifx [1]{%
 \ifx #1\expandafter \@firstoftwo
 \else \expandafter \@secondoftwo
 \fi
}%
\providecommand \natexlab [1]{#1}%
\providecommand \enquote  [1]{``#1''}%
\providecommand \bibnamefont  [1]{#1}%
\providecommand \bibfnamefont [1]{#1}%
\providecommand \citenamefont [1]{#1}%
\providecommand \href@noop [0]{\@secondoftwo}%
\providecommand \href [0]{\begingroup \@sanitize@url \@href}%
\providecommand \@href[1]{\@@startlink{#1}\@@href}%
\providecommand \@@href[1]{\endgroup#1\@@endlink}%
\providecommand \@sanitize@url [0]{\catcode `\\12\catcode `\$12\catcode
  `\&12\catcode `\#12\catcode `\^12\catcode `\_12\catcode `\%12\relax}%
\providecommand \@@startlink[1]{}%
\providecommand \@@endlink[0]{}%
\providecommand \url  [0]{\begingroup\@sanitize@url \@url }%
\providecommand \@url [1]{\endgroup\@href {#1}{\urlprefix }}%
\providecommand \urlprefix  [0]{URL }%
\providecommand \Eprint [0]{\href }%
\providecommand \doibase [0]{http://dx.doi.org/}%
\providecommand \selectlanguage [0]{\@gobble}%
\providecommand \bibinfo  [0]{\@secondoftwo}%
\providecommand \bibfield  [0]{\@secondoftwo}%
\providecommand \translation [1]{[#1]}%
\providecommand \BibitemOpen [0]{}%
\providecommand \bibitemStop [0]{}%
\providecommand \bibitemNoStop [0]{.\EOS\space}%
\providecommand \EOS [0]{\spacefactor3000\relax}%
\providecommand \BibitemShut  [1]{\csname bibitem#1\endcsname}%
\let\auto@bib@innerbib\@empty
\bibitem [{\citenamefont {Mazurenko}\ \emph {et~al.}(2017)\citenamefont
  {Mazurenko}, \citenamefont {Chiu}, \citenamefont {Ji}, \citenamefont
  {Parsons}, \citenamefont {Kan{\'a}sz-Nagy}, \citenamefont {Schmidt},
  \citenamefont {Grusdt}, \citenamefont {Demler}, \citenamefont {Greif},\ and\
  \citenamefont {Greiner}}]{mazurenko2017cold}%
  \BibitemOpen
  \bibfield  {author} {\bibinfo {author} {\bibfnamefont {A.}~\bibnamefont
  {Mazurenko}}, \bibinfo {author} {\bibfnamefont {C.~S.}\ \bibnamefont {Chiu}},
  \bibinfo {author} {\bibfnamefont {G.}~\bibnamefont {Ji}}, \bibinfo {author}
  {\bibfnamefont {M.~F.}\ \bibnamefont {Parsons}}, \bibinfo {author}
  {\bibfnamefont {M.}~\bibnamefont {Kan{\'a}sz-Nagy}}, \bibinfo {author}
  {\bibfnamefont {R.}~\bibnamefont {Schmidt}}, \bibinfo {author} {\bibfnamefont
  {F.}~\bibnamefont {Grusdt}}, \bibinfo {author} {\bibfnamefont
  {E.}~\bibnamefont {Demler}}, \bibinfo {author} {\bibfnamefont
  {D.}~\bibnamefont {Greif}}, \ and\ \bibinfo {author} {\bibfnamefont
  {M.}~\bibnamefont {Greiner}},\ }\href@noop {} {\bibfield  {journal} {\bibinfo
   {journal} {Nature}\ }\textbf {\bibinfo {volume} {545}},\ \bibinfo {pages}
  {462} (\bibinfo {year} {2017})}\BibitemShut {NoStop}%
\bibitem [{\citenamefont {Cheuk}\ \emph {et~al.}(2016)\citenamefont {Cheuk},
  \citenamefont {Nichols}, \citenamefont {Lawrence}, \citenamefont {Okan},
  \citenamefont {Zhang}, \citenamefont {Khatami}, \citenamefont {Trivedi},
  \citenamefont {Paiva}, \citenamefont {Rigol},\ and\ \citenamefont
  {Zwierlein}}]{cheuk2016observation}%
  \BibitemOpen
  \bibfield  {author} {\bibinfo {author} {\bibfnamefont {L.~W.}\ \bibnamefont
  {Cheuk}}, \bibinfo {author} {\bibfnamefont {M.~A.}\ \bibnamefont {Nichols}},
  \bibinfo {author} {\bibfnamefont {K.~R.}\ \bibnamefont {Lawrence}}, \bibinfo
  {author} {\bibfnamefont {M.}~\bibnamefont {Okan}}, \bibinfo {author}
  {\bibfnamefont {H.}~\bibnamefont {Zhang}}, \bibinfo {author} {\bibfnamefont
  {E.}~\bibnamefont {Khatami}}, \bibinfo {author} {\bibfnamefont
  {N.}~\bibnamefont {Trivedi}}, \bibinfo {author} {\bibfnamefont
  {T.}~\bibnamefont {Paiva}}, \bibinfo {author} {\bibfnamefont
  {M.}~\bibnamefont {Rigol}}, \ and\ \bibinfo {author} {\bibfnamefont {M.~W.}\
  \bibnamefont {Zwierlein}},\ }\href@noop {} {\bibfield  {journal} {\bibinfo
  {journal} {Science}\ }\textbf {\bibinfo {volume} {353}},\ \bibinfo {pages}
  {1260} (\bibinfo {year} {2016})}\BibitemShut {NoStop}%
\bibitem [{\citenamefont {L\"uschen}\ \emph {et~al.}(2017)\citenamefont
  {L\"uschen}, \citenamefont {Bordia}, \citenamefont {Hodgman}, \citenamefont
  {Schreiber}, \citenamefont {Sarkar}, \citenamefont {Daley}, \citenamefont
  {Fischer}, \citenamefont {Altman}, \citenamefont {Bloch},\ and\ \citenamefont
  {Schneider}}]{schneiderMBL}%
  \BibitemOpen
  \bibfield  {author} {\bibinfo {author} {\bibfnamefont {H.~P.}\ \bibnamefont
  {L\"uschen}}, \bibinfo {author} {\bibfnamefont {P.}~\bibnamefont {Bordia}},
  \bibinfo {author} {\bibfnamefont {S.~S.}\ \bibnamefont {Hodgman}}, \bibinfo
  {author} {\bibfnamefont {M.}~\bibnamefont {Schreiber}}, \bibinfo {author}
  {\bibfnamefont {S.}~\bibnamefont {Sarkar}}, \bibinfo {author} {\bibfnamefont
  {A.~J.}\ \bibnamefont {Daley}}, \bibinfo {author} {\bibfnamefont {M.~H.}\
  \bibnamefont {Fischer}}, \bibinfo {author} {\bibfnamefont {E.}~\bibnamefont
  {Altman}}, \bibinfo {author} {\bibfnamefont {I.}~\bibnamefont {Bloch}}, \
  and\ \bibinfo {author} {\bibfnamefont {U.}~\bibnamefont {Schneider}},\ }\href
  {\doibase 10.1103/PhysRevX.7.011034} {\bibfield  {journal} {\bibinfo
  {journal} {Phys. Rev. X}\ }\textbf {\bibinfo {volume} {7}},\ \bibinfo {pages}
  {011034} (\bibinfo {year} {2017})}\BibitemShut {NoStop}%
\bibitem [{\citenamefont {L{\"u}schen}\ \emph {et~al.}(2016)\citenamefont
  {L{\"u}schen}, \citenamefont {Bordia}, \citenamefont {Scherg}, \citenamefont
  {Alet}, \citenamefont {Altman}, \citenamefont {Schneider},\ and\
  \citenamefont {Bloch}}]{luschen2016evidence}%
  \BibitemOpen
  \bibfield  {author} {\bibinfo {author} {\bibfnamefont {H.~P.}\ \bibnamefont
  {L{\"u}schen}}, \bibinfo {author} {\bibfnamefont {P.}~\bibnamefont {Bordia}},
  \bibinfo {author} {\bibfnamefont {S.}~\bibnamefont {Scherg}}, \bibinfo
  {author} {\bibfnamefont {F.}~\bibnamefont {Alet}}, \bibinfo {author}
  {\bibfnamefont {E.}~\bibnamefont {Altman}}, \bibinfo {author} {\bibfnamefont
  {U.}~\bibnamefont {Schneider}}, \ and\ \bibinfo {author} {\bibfnamefont
  {I.}~\bibnamefont {Bloch}},\ }\href@noop {} {\bibfield  {journal} {\bibinfo
  {journal} {arXiv preprint arXiv:1612.07173}\ } (\bibinfo {year}
  {2016})}\BibitemShut {NoStop}%
\bibitem [{\citenamefont {Langen}\ \emph {et~al.}(2015)\citenamefont {Langen},
  \citenamefont {Geiger},\ and\ \citenamefont
  {Schmiedmayer}}]{langen2015ultracold}%
  \BibitemOpen
  \bibfield  {author} {\bibinfo {author} {\bibfnamefont {T.}~\bibnamefont
  {Langen}}, \bibinfo {author} {\bibfnamefont {R.}~\bibnamefont {Geiger}}, \
  and\ \bibinfo {author} {\bibfnamefont {J.}~\bibnamefont {Schmiedmayer}},\
  }\href@noop {} {\  (\bibinfo {year} {2015})}\BibitemShut {NoStop}%
\bibitem [{\citenamefont {Langen}\ \emph {et~al.}(2013)\citenamefont {Langen},
  \citenamefont {Geiger}, \citenamefont {Kuhnert}, \citenamefont {Rauer},\ and\
  \citenamefont {Schmiedmayer}}]{Langen}%
  \BibitemOpen
  \bibfield  {author} {\bibinfo {author} {\bibfnamefont {T.}~\bibnamefont
  {Langen}}, \bibinfo {author} {\bibfnamefont {R.}~\bibnamefont {Geiger}},
  \bibinfo {author} {\bibfnamefont {M.}~\bibnamefont {Kuhnert}}, \bibinfo
  {author} {\bibfnamefont {B.}~\bibnamefont {Rauer}}, \ and\ \bibinfo {author}
  {\bibfnamefont {J.}~\bibnamefont {Schmiedmayer}},\ }\href@noop {} {\bibfield
  {journal} {\bibinfo  {journal} {Nature Physics}\ }\textbf {\bibinfo {volume}
  {9}},\ \bibinfo {pages} {640} (\bibinfo {year} {2013})}\BibitemShut {NoStop}%
\bibitem [{\citenamefont {Fuchs}\ \emph {et~al.}(2011)\citenamefont {Fuchs},
  \citenamefont {Burkard}, \citenamefont {Klimov},\ and\ \citenamefont
  {Awschalom}}]{fuchsNV}%
  \BibitemOpen
  \bibfield  {author} {\bibinfo {author} {\bibfnamefont {G.}~\bibnamefont
  {Fuchs}}, \bibinfo {author} {\bibfnamefont {G.}~\bibnamefont {Burkard}},
  \bibinfo {author} {\bibfnamefont {P.}~\bibnamefont {Klimov}}, \ and\ \bibinfo
  {author} {\bibfnamefont {D.}~\bibnamefont {Awschalom}},\ }\href
  {http://www.nature.com/nphys/journal/v7/n10/abs/nphys2026.html} {\bibfield
  {journal} {\bibinfo  {journal} {Nature Physics}\ }\textbf {\bibinfo {volume}
  {7}},\ \bibinfo {pages} {789} (\bibinfo {year} {2011})}\BibitemShut {NoStop}%
\bibitem [{\citenamefont {Dutt}\ \emph {et~al.}(2007)\citenamefont {Dutt},
  \citenamefont {Childress}, \citenamefont {Jiang}, \citenamefont {Togan},
  \citenamefont {Maze}, \citenamefont {Jelezko}, \citenamefont {Zibrov},
  \citenamefont {Hemmer},\ and\ \citenamefont {Lukin}}]{DuttNV}%
  \BibitemOpen
  \bibfield  {author} {\bibinfo {author} {\bibfnamefont {M.~V.~G.}\
  \bibnamefont {Dutt}}, \bibinfo {author} {\bibfnamefont {L.}~\bibnamefont
  {Childress}}, \bibinfo {author} {\bibfnamefont {L.}~\bibnamefont {Jiang}},
  \bibinfo {author} {\bibfnamefont {E.}~\bibnamefont {Togan}}, \bibinfo
  {author} {\bibfnamefont {J.}~\bibnamefont {Maze}}, \bibinfo {author}
  {\bibfnamefont {F.}~\bibnamefont {Jelezko}}, \bibinfo {author} {\bibfnamefont
  {A.~S.}\ \bibnamefont {Zibrov}}, \bibinfo {author} {\bibfnamefont {P.~R.}\
  \bibnamefont {Hemmer}}, \ and\ \bibinfo {author} {\bibfnamefont {M.~D.}\
  \bibnamefont {Lukin}},\ }\href {\doibase 10.1126/science.1139831} {\bibfield
  {journal} {\bibinfo  {journal} {Science}\ }\textbf {\bibinfo {volume}
  {316}},\ \bibinfo {pages} {1312} (\bibinfo {year} {2007})}\BibitemShut
  {NoStop}%
\bibitem [{\citenamefont {Maurer}\ \emph {et~al.}(2012)\citenamefont {Maurer},
  \citenamefont {Kucsko}, \citenamefont {Latta}, \citenamefont {Jiang},
  \citenamefont {Yao}, \citenamefont {Bennett}, \citenamefont {Pastawski},
  \citenamefont {Hunger}, \citenamefont {Chisholm}, \citenamefont {Markham},
  \citenamefont {Twitchen}, \citenamefont {Cirac},\ and\ \citenamefont
  {Lukin}}]{MaurerNV}%
  \BibitemOpen
  \bibfield  {author} {\bibinfo {author} {\bibfnamefont {P.~C.}\ \bibnamefont
  {Maurer}}, \bibinfo {author} {\bibfnamefont {G.}~\bibnamefont {Kucsko}},
  \bibinfo {author} {\bibfnamefont {C.}~\bibnamefont {Latta}}, \bibinfo
  {author} {\bibfnamefont {L.}~\bibnamefont {Jiang}}, \bibinfo {author}
  {\bibfnamefont {N.~Y.}\ \bibnamefont {Yao}}, \bibinfo {author} {\bibfnamefont
  {S.~D.}\ \bibnamefont {Bennett}}, \bibinfo {author} {\bibfnamefont
  {F.}~\bibnamefont {Pastawski}}, \bibinfo {author} {\bibfnamefont
  {D.}~\bibnamefont {Hunger}}, \bibinfo {author} {\bibfnamefont
  {N.}~\bibnamefont {Chisholm}}, \bibinfo {author} {\bibfnamefont
  {M.}~\bibnamefont {Markham}}, \bibinfo {author} {\bibfnamefont {D.~J.}\
  \bibnamefont {Twitchen}}, \bibinfo {author} {\bibfnamefont {J.~I.}\
  \bibnamefont {Cirac}}, \ and\ \bibinfo {author} {\bibfnamefont {M.~D.}\
  \bibnamefont {Lukin}},\ }\href {\doibase 10.1126/science.1220513} {\bibfield
  {journal} {\bibinfo  {journal} {Science}\ }\textbf {\bibinfo {volume}
  {336}},\ \bibinfo {pages} {1283} (\bibinfo {year} {2012})}\BibitemShut
  {NoStop}%
\bibitem [{\citenamefont {Sushkov}\ \emph {et~al.}(2014)\citenamefont
  {Sushkov}, \citenamefont {Lovchinsky}, \citenamefont {Chisholm},
  \citenamefont {Walsworth}, \citenamefont {Park},\ and\ \citenamefont
  {Lukin}}]{ShuskovNV}%
  \BibitemOpen
  \bibfield  {author} {\bibinfo {author} {\bibfnamefont {A.~O.}\ \bibnamefont
  {Sushkov}}, \bibinfo {author} {\bibfnamefont {I.}~\bibnamefont {Lovchinsky}},
  \bibinfo {author} {\bibfnamefont {N.}~\bibnamefont {Chisholm}}, \bibinfo
  {author} {\bibfnamefont {R.~L.}\ \bibnamefont {Walsworth}}, \bibinfo {author}
  {\bibfnamefont {H.}~\bibnamefont {Park}}, \ and\ \bibinfo {author}
  {\bibfnamefont {M.~D.}\ \bibnamefont {Lukin}},\ }\href {\doibase
  10.1103/PhysRevLett.113.197601} {\bibfield  {journal} {\bibinfo  {journal}
  {Phys. Rev. Lett.}\ }\textbf {\bibinfo {volume} {113}},\ \bibinfo {pages}
  {197601} (\bibinfo {year} {2014})}\BibitemShut {NoStop}%
\bibitem [{\citenamefont {Agarwal}\ \emph
  {et~al.}(2017{\natexlab{a}})\citenamefont {Agarwal}, \citenamefont {Schmidt},
  \citenamefont {Halperin}, \citenamefont {Oganesyan}, \citenamefont
  {Zar\'and}, \citenamefont {Lukin},\ and\ \citenamefont
  {Demler}}]{AgarwalmagneticnoiseNV}%
  \BibitemOpen
  \bibfield  {author} {\bibinfo {author} {\bibfnamefont {K.}~\bibnamefont
  {Agarwal}}, \bibinfo {author} {\bibfnamefont {R.}~\bibnamefont {Schmidt}},
  \bibinfo {author} {\bibfnamefont {B.}~\bibnamefont {Halperin}}, \bibinfo
  {author} {\bibfnamefont {V.}~\bibnamefont {Oganesyan}}, \bibinfo {author}
  {\bibfnamefont {G.}~\bibnamefont {Zar\'and}}, \bibinfo {author}
  {\bibfnamefont {M.~D.}\ \bibnamefont {Lukin}}, \ and\ \bibinfo {author}
  {\bibfnamefont {E.}~\bibnamefont {Demler}},\ }\href {\doibase
  10.1103/PhysRevB.95.155107} {\bibfield  {journal} {\bibinfo  {journal} {Phys.
  Rev. B}\ }\textbf {\bibinfo {volume} {95}},\ \bibinfo {pages} {155107}
  (\bibinfo {year} {2017}{\natexlab{a}})}\BibitemShut {NoStop}%
\bibitem [{\citenamefont {Choi}\ \emph {et~al.}(2017)\citenamefont {Choi},
  \citenamefont {Choi}, \citenamefont {Landig}, \citenamefont {Kucsko},
  \citenamefont {Zhou}, \citenamefont {Isoya}, \citenamefont {Jelezko},
  \citenamefont {Onoda}, \citenamefont {Sumiya}, \citenamefont {Khemani} \emph
  {et~al.}}]{choi2017observation}%
  \BibitemOpen
  \bibfield  {author} {\bibinfo {author} {\bibfnamefont {S.}~\bibnamefont
  {Choi}}, \bibinfo {author} {\bibfnamefont {J.}~\bibnamefont {Choi}}, \bibinfo
  {author} {\bibfnamefont {R.}~\bibnamefont {Landig}}, \bibinfo {author}
  {\bibfnamefont {G.}~\bibnamefont {Kucsko}}, \bibinfo {author} {\bibfnamefont
  {H.}~\bibnamefont {Zhou}}, \bibinfo {author} {\bibfnamefont {J.}~\bibnamefont
  {Isoya}}, \bibinfo {author} {\bibfnamefont {F.}~\bibnamefont {Jelezko}},
  \bibinfo {author} {\bibfnamefont {S.}~\bibnamefont {Onoda}}, \bibinfo
  {author} {\bibfnamefont {H.}~\bibnamefont {Sumiya}}, \bibinfo {author}
  {\bibfnamefont {V.}~\bibnamefont {Khemani}},  \emph {et~al.},\ }\href@noop {}
  {\bibfield  {journal} {\bibinfo  {journal} {Nature}\ }\textbf {\bibinfo
  {volume} {543}},\ \bibinfo {pages} {221} (\bibinfo {year}
  {2017})}\BibitemShut {NoStop}%
\bibitem [{\citenamefont {Debnath}\ \emph {et~al.}(2016)\citenamefont
  {Debnath}, \citenamefont {Linke}, \citenamefont {Figgatt}, \citenamefont
  {Landsman}, \citenamefont {Wright},\ and\ \citenamefont
  {Monroe}}]{debnath2016demonstration}%
  \BibitemOpen
  \bibfield  {author} {\bibinfo {author} {\bibfnamefont {S.}~\bibnamefont
  {Debnath}}, \bibinfo {author} {\bibfnamefont {N.}~\bibnamefont {Linke}},
  \bibinfo {author} {\bibfnamefont {C.}~\bibnamefont {Figgatt}}, \bibinfo
  {author} {\bibfnamefont {K.}~\bibnamefont {Landsman}}, \bibinfo {author}
  {\bibfnamefont {K.}~\bibnamefont {Wright}}, \ and\ \bibinfo {author}
  {\bibfnamefont {C.}~\bibnamefont {Monroe}},\ }\href@noop {} {\bibfield
  {journal} {\bibinfo  {journal} {Nature}\ }\textbf {\bibinfo {volume} {536}},\
  \bibinfo {pages} {63} (\bibinfo {year} {2016})}\BibitemShut {NoStop}%
\bibitem [{\citenamefont {Zhang}\ \emph {et~al.}(2017)\citenamefont {Zhang},
  \citenamefont {Pagano}, \citenamefont {Hess}, \citenamefont {Kyprianidis},
  \citenamefont {Becker}, \citenamefont {Kaplan}, \citenamefont {Gorshkov},
  \citenamefont {Gong},\ and\ \citenamefont {Monroe}}]{zhang2017observation}%
  \BibitemOpen
  \bibfield  {author} {\bibinfo {author} {\bibfnamefont {J.}~\bibnamefont
  {Zhang}}, \bibinfo {author} {\bibfnamefont {G.}~\bibnamefont {Pagano}},
  \bibinfo {author} {\bibfnamefont {P.}~\bibnamefont {Hess}}, \bibinfo {author}
  {\bibfnamefont {A.}~\bibnamefont {Kyprianidis}}, \bibinfo {author}
  {\bibfnamefont {P.}~\bibnamefont {Becker}}, \bibinfo {author} {\bibfnamefont
  {H.}~\bibnamefont {Kaplan}}, \bibinfo {author} {\bibfnamefont
  {A.}~\bibnamefont {Gorshkov}}, \bibinfo {author} {\bibfnamefont {Z.-X.}\
  \bibnamefont {Gong}}, \ and\ \bibinfo {author} {\bibfnamefont
  {C.}~\bibnamefont {Monroe}},\ }\href@noop {} {\bibfield  {journal} {\bibinfo
  {journal} {arXiv preprint arXiv:1708.01044}\ } (\bibinfo {year}
  {2017})}\BibitemShut {NoStop}%
\bibitem [{\citenamefont {Jurcevic}\ \emph {et~al.}(2014)\citenamefont
  {Jurcevic}, \citenamefont {Lanyon}, \citenamefont {Hauke}, \citenamefont
  {Hempel}, \citenamefont {Zoller}, \citenamefont {Blatt},\ and\ \citenamefont
  {Roos}}]{jurcevic2014observation}%
  \BibitemOpen
  \bibfield  {author} {\bibinfo {author} {\bibfnamefont {P.}~\bibnamefont
  {Jurcevic}}, \bibinfo {author} {\bibfnamefont {B.~P.}\ \bibnamefont
  {Lanyon}}, \bibinfo {author} {\bibfnamefont {P.}~\bibnamefont {Hauke}},
  \bibinfo {author} {\bibfnamefont {C.}~\bibnamefont {Hempel}}, \bibinfo
  {author} {\bibfnamefont {P.}~\bibnamefont {Zoller}}, \bibinfo {author}
  {\bibfnamefont {R.}~\bibnamefont {Blatt}}, \ and\ \bibinfo {author}
  {\bibfnamefont {C.~F.}\ \bibnamefont {Roos}},\ }\href@noop {} {\bibfield
  {journal} {\bibinfo  {journal} {arXiv preprint arXiv:1401.5387}\ } (\bibinfo
  {year} {2014})}\BibitemShut {NoStop}%
\bibitem [{\citenamefont {Barends}\ \emph {et~al.}(2016)\citenamefont
  {Barends}, \citenamefont {Shabani}, \citenamefont {Lamata}, \citenamefont
  {Kelly}, \citenamefont {Mezzacapo}, \citenamefont {Las~Heras}, \citenamefont
  {Babbush}, \citenamefont {Fowler}, \citenamefont {Campbell}, \citenamefont
  {Chen} \emph {et~al.}}]{barends2016digitized}%
  \BibitemOpen
  \bibfield  {author} {\bibinfo {author} {\bibfnamefont {R.}~\bibnamefont
  {Barends}}, \bibinfo {author} {\bibfnamefont {A.}~\bibnamefont {Shabani}},
  \bibinfo {author} {\bibfnamefont {L.}~\bibnamefont {Lamata}}, \bibinfo
  {author} {\bibfnamefont {J.}~\bibnamefont {Kelly}}, \bibinfo {author}
  {\bibfnamefont {A.}~\bibnamefont {Mezzacapo}}, \bibinfo {author}
  {\bibfnamefont {U.}~\bibnamefont {Las~Heras}}, \bibinfo {author}
  {\bibfnamefont {R.}~\bibnamefont {Babbush}}, \bibinfo {author} {\bibfnamefont
  {A.~G.}\ \bibnamefont {Fowler}}, \bibinfo {author} {\bibfnamefont
  {B.}~\bibnamefont {Campbell}}, \bibinfo {author} {\bibfnamefont
  {Y.}~\bibnamefont {Chen}},  \emph {et~al.},\ }\href@noop {} {\bibfield
  {journal} {\bibinfo  {journal} {Nature}\ }\textbf {\bibinfo {volume} {534}},\
  \bibinfo {pages} {222} (\bibinfo {year} {2016})}\BibitemShut {NoStop}%
\bibitem [{\citenamefont {Boixo}\ \emph {et~al.}(2013)\citenamefont {Boixo},
  \citenamefont {R{\o}nnow}, \citenamefont {Isakov}, \citenamefont {Wang},
  \citenamefont {Wecker}, \citenamefont {Lidar}, \citenamefont {Martinis},\
  and\ \citenamefont {Troyer}}]{boixo2013quantum}%
  \BibitemOpen
  \bibfield  {author} {\bibinfo {author} {\bibfnamefont {S.}~\bibnamefont
  {Boixo}}, \bibinfo {author} {\bibfnamefont {T.~F.}\ \bibnamefont
  {R{\o}nnow}}, \bibinfo {author} {\bibfnamefont {S.~V.}\ \bibnamefont
  {Isakov}}, \bibinfo {author} {\bibfnamefont {Z.}~\bibnamefont {Wang}},
  \bibinfo {author} {\bibfnamefont {D.}~\bibnamefont {Wecker}}, \bibinfo
  {author} {\bibfnamefont {D.~A.}\ \bibnamefont {Lidar}}, \bibinfo {author}
  {\bibfnamefont {J.~M.}\ \bibnamefont {Martinis}}, \ and\ \bibinfo {author}
  {\bibfnamefont {M.}~\bibnamefont {Troyer}},\ }\href@noop {} {\bibfield
  {journal} {\bibinfo  {journal} {arXiv preprint arXiv:1304.4595}\ } (\bibinfo
  {year} {2013})}\BibitemShut {NoStop}%
\bibitem [{\citenamefont {Johnson}\ \emph {et~al.}(2011)\citenamefont
  {Johnson}, \citenamefont {Amin}, \citenamefont {Gildert}, \citenamefont
  {Lanting}, \citenamefont {Hamze}, \citenamefont {Dickson}, \citenamefont
  {Harris}, \citenamefont {Berkley}, \citenamefont {Johansson}, \citenamefont
  {Bunyk} \emph {et~al.}}]{johnson2011quantum}%
  \BibitemOpen
  \bibfield  {author} {\bibinfo {author} {\bibfnamefont {M.~W.}\ \bibnamefont
  {Johnson}}, \bibinfo {author} {\bibfnamefont {M.~H.}\ \bibnamefont {Amin}},
  \bibinfo {author} {\bibfnamefont {S.}~\bibnamefont {Gildert}}, \bibinfo
  {author} {\bibfnamefont {T.}~\bibnamefont {Lanting}}, \bibinfo {author}
  {\bibfnamefont {F.}~\bibnamefont {Hamze}}, \bibinfo {author} {\bibfnamefont
  {N.}~\bibnamefont {Dickson}}, \bibinfo {author} {\bibfnamefont
  {R.}~\bibnamefont {Harris}}, \bibinfo {author} {\bibfnamefont {A.~J.}\
  \bibnamefont {Berkley}}, \bibinfo {author} {\bibfnamefont {J.}~\bibnamefont
  {Johansson}}, \bibinfo {author} {\bibfnamefont {P.}~\bibnamefont {Bunyk}},
  \emph {et~al.},\ }\href@noop {} {\bibfield  {journal} {\bibinfo  {journal}
  {Nature}\ }\textbf {\bibinfo {volume} {473}},\ \bibinfo {pages} {194}
  (\bibinfo {year} {2011})}\BibitemShut {NoStop}%
\bibitem [{\citenamefont {Lanting}\ \emph {et~al.}(2014)\citenamefont
  {Lanting}, \citenamefont {Przybysz}, \citenamefont {Smirnov}, \citenamefont
  {Spedalieri}, \citenamefont {Amin}, \citenamefont {Berkley}, \citenamefont
  {Harris}, \citenamefont {Altomare}, \citenamefont {Boixo}, \citenamefont
  {Bunyk}, \citenamefont {Dickson}, \citenamefont {Enderud}, \citenamefont
  {Hilton}, \citenamefont {Hoskinson}, \citenamefont {Johnson}, \citenamefont
  {Ladizinsky}, \citenamefont {Ladizinsky}, \citenamefont {Neufeld},
  \citenamefont {Oh}, \citenamefont {Perminov}, \citenamefont {Rich},
  \citenamefont {Thom}, \citenamefont {Tolkacheva}, \citenamefont {Uchaikin},
  \citenamefont {Wilson},\ and\ \citenamefont {Rose}}]{dwaveannealing}%
  \BibitemOpen
  \bibfield  {author} {\bibinfo {author} {\bibfnamefont {T.}~\bibnamefont
  {Lanting}}, \bibinfo {author} {\bibfnamefont {A.~J.}\ \bibnamefont
  {Przybysz}}, \bibinfo {author} {\bibfnamefont {A.~Y.}\ \bibnamefont
  {Smirnov}}, \bibinfo {author} {\bibfnamefont {F.~M.}\ \bibnamefont
  {Spedalieri}}, \bibinfo {author} {\bibfnamefont {M.~H.}\ \bibnamefont
  {Amin}}, \bibinfo {author} {\bibfnamefont {A.~J.}\ \bibnamefont {Berkley}},
  \bibinfo {author} {\bibfnamefont {R.}~\bibnamefont {Harris}}, \bibinfo
  {author} {\bibfnamefont {F.}~\bibnamefont {Altomare}}, \bibinfo {author}
  {\bibfnamefont {S.}~\bibnamefont {Boixo}}, \bibinfo {author} {\bibfnamefont
  {P.}~\bibnamefont {Bunyk}}, \bibinfo {author} {\bibfnamefont
  {N.}~\bibnamefont {Dickson}}, \bibinfo {author} {\bibfnamefont
  {C.}~\bibnamefont {Enderud}}, \bibinfo {author} {\bibfnamefont {J.~P.}\
  \bibnamefont {Hilton}}, \bibinfo {author} {\bibfnamefont {E.}~\bibnamefont
  {Hoskinson}}, \bibinfo {author} {\bibfnamefont {M.~W.}\ \bibnamefont
  {Johnson}}, \bibinfo {author} {\bibfnamefont {E.}~\bibnamefont {Ladizinsky}},
  \bibinfo {author} {\bibfnamefont {N.}~\bibnamefont {Ladizinsky}}, \bibinfo
  {author} {\bibfnamefont {R.}~\bibnamefont {Neufeld}}, \bibinfo {author}
  {\bibfnamefont {T.}~\bibnamefont {Oh}}, \bibinfo {author} {\bibfnamefont
  {I.}~\bibnamefont {Perminov}}, \bibinfo {author} {\bibfnamefont
  {C.}~\bibnamefont {Rich}}, \bibinfo {author} {\bibfnamefont {M.~C.}\
  \bibnamefont {Thom}}, \bibinfo {author} {\bibfnamefont {E.}~\bibnamefont
  {Tolkacheva}}, \bibinfo {author} {\bibfnamefont {S.}~\bibnamefont
  {Uchaikin}}, \bibinfo {author} {\bibfnamefont {A.~B.}\ \bibnamefont
  {Wilson}}, \ and\ \bibinfo {author} {\bibfnamefont {G.}~\bibnamefont
  {Rose}},\ }\href {\doibase 10.1103/PhysRevX.4.021041} {\bibfield  {journal}
  {\bibinfo  {journal} {Phys. Rev. X}\ }\textbf {\bibinfo {volume} {4}},\
  \bibinfo {pages} {021041} (\bibinfo {year} {2014})}\BibitemShut {NoStop}%
\bibitem [{\citenamefont {Griffiths}\ and\ \citenamefont
  {Harris}(1995)}]{griffiths1995introduction}%
  \BibitemOpen
  \bibfield  {author} {\bibinfo {author} {\bibfnamefont {D.~J.}\ \bibnamefont
  {Griffiths}}\ and\ \bibinfo {author} {\bibfnamefont {E.~G.}\ \bibnamefont
  {Harris}},\ }\href@noop {} {\bibfield  {journal} {\bibinfo  {journal}
  {American Journal of Physics}\ }\textbf {\bibinfo {volume} {63}},\ \bibinfo
  {pages} {767} (\bibinfo {year} {1995})}\BibitemShut {NoStop}%
\bibitem [{\citenamefont {Young}\ \emph {et~al.}(2008)\citenamefont {Young},
  \citenamefont {Knysh},\ and\ \citenamefont {Smelyanskiy}}]{mingapYoung}%
  \BibitemOpen
  \bibfield  {author} {\bibinfo {author} {\bibfnamefont {A.~P.}\ \bibnamefont
  {Young}}, \bibinfo {author} {\bibfnamefont {S.}~\bibnamefont {Knysh}}, \ and\
  \bibinfo {author} {\bibfnamefont {V.~N.}\ \bibnamefont {Smelyanskiy}},\
  }\href {\doibase 10.1103/PhysRevLett.101.170503} {\bibfield  {journal}
  {\bibinfo  {journal} {Phys. Rev. Lett.}\ }\textbf {\bibinfo {volume} {101}},\
  \bibinfo {pages} {170503} (\bibinfo {year} {2008})}\BibitemShut {NoStop}%
\bibitem [{\citenamefont {Hen}\ and\ \citenamefont
  {Young}(2011)}]{QAAsatisfiability}%
  \BibitemOpen
  \bibfield  {author} {\bibinfo {author} {\bibfnamefont {I.}~\bibnamefont
  {Hen}}\ and\ \bibinfo {author} {\bibfnamefont {A.~P.}\ \bibnamefont
  {Young}},\ }\href {\doibase 10.1103/PhysRevE.84.061152} {\bibfield  {journal}
  {\bibinfo  {journal} {Phys. Rev. E}\ }\textbf {\bibinfo {volume} {84}},\
  \bibinfo {pages} {061152} (\bibinfo {year} {2011})}\BibitemShut {NoStop}%
\bibitem [{\citenamefont {Farhi}\ \emph {et~al.}(2012)\citenamefont {Farhi},
  \citenamefont {Gosset}, \citenamefont {Hen}, \citenamefont {Sandvik},
  \citenamefont {Shor}, \citenamefont {Young},\ and\ \citenamefont
  {Zamponi}}]{QAAparticular}%
  \BibitemOpen
  \bibfield  {author} {\bibinfo {author} {\bibfnamefont {E.}~\bibnamefont
  {Farhi}}, \bibinfo {author} {\bibfnamefont {D.}~\bibnamefont {Gosset}},
  \bibinfo {author} {\bibfnamefont {I.}~\bibnamefont {Hen}}, \bibinfo {author}
  {\bibfnamefont {A.~W.}\ \bibnamefont {Sandvik}}, \bibinfo {author}
  {\bibfnamefont {P.}~\bibnamefont {Shor}}, \bibinfo {author} {\bibfnamefont
  {A.~P.}\ \bibnamefont {Young}}, \ and\ \bibinfo {author} {\bibfnamefont
  {F.}~\bibnamefont {Zamponi}},\ }\href {\doibase 10.1103/PhysRevA.86.052334}
  {\bibfield  {journal} {\bibinfo  {journal} {Phys. Rev. A}\ }\textbf {\bibinfo
  {volume} {86}},\ \bibinfo {pages} {052334} (\bibinfo {year}
  {2012})}\BibitemShut {NoStop}%
\bibitem [{\citenamefont {Agarwal}\ \emph {et~al.}(2018)\citenamefont
  {Agarwal}, \citenamefont {Bhatt},\ and\ \citenamefont
  {Sondhi}}]{AgarwalGroundStatePrep}%
  \BibitemOpen
  \bibfield  {author} {\bibinfo {author} {\bibfnamefont {K.}~\bibnamefont
  {Agarwal}}, \bibinfo {author} {\bibfnamefont {R.~N.}\ \bibnamefont {Bhatt}},
  \ and\ \bibinfo {author} {\bibfnamefont {S.~L.}\ \bibnamefont {Sondhi}},\
  }\href {\doibase 10.1103/PhysRevLett.120.210604} {\bibfield  {journal}
  {\bibinfo  {journal} {Phys. Rev. Lett.}\ }\textbf {\bibinfo {volume} {120}},\
  \bibinfo {pages} {210604} (\bibinfo {year} {2018})}\BibitemShut {NoStop}%
\bibitem [{\citenamefont {del
  Campo}(2013{\natexlab{a}})}]{delCampoCounterDiabatic}%
  \BibitemOpen
  \bibfield  {author} {\bibinfo {author} {\bibfnamefont {A.}~\bibnamefont {del
  Campo}},\ }\href {\doibase 10.1103/PhysRevLett.111.100502} {\bibfield
  {journal} {\bibinfo  {journal} {Phys. Rev. Lett.}\ }\textbf {\bibinfo
  {volume} {111}},\ \bibinfo {pages} {100502} (\bibinfo {year}
  {2013}{\natexlab{a}})}\BibitemShut {NoStop}%
\bibitem [{\citenamefont {del Campo}(2013{\natexlab{b}})}]{del2013shortcuts}%
  \BibitemOpen
  \bibfield  {author} {\bibinfo {author} {\bibfnamefont {A.}~\bibnamefont {del
  Campo}},\ }\href@noop {} {\bibfield  {journal} {\bibinfo  {journal} {Physical
  review letters}\ }\textbf {\bibinfo {volume} {111}},\ \bibinfo {pages}
  {100502} (\bibinfo {year} {2013}{\natexlab{b}})}\BibitemShut {NoStop}%
\bibitem [{\citenamefont {Jarzynski}(2013)}]{jarzynskitransitionless}%
  \BibitemOpen
  \bibfield  {author} {\bibinfo {author} {\bibfnamefont {C.}~\bibnamefont
  {Jarzynski}},\ }\href {\doibase 10.1103/PhysRevA.88.040101} {\bibfield
  {journal} {\bibinfo  {journal} {Phys. Rev. A}\ }\textbf {\bibinfo {volume}
  {88}},\ \bibinfo {pages} {040101} (\bibinfo {year} {2013})}\BibitemShut
  {NoStop}%
\bibitem [{\citenamefont {Glaser}\ \emph {et~al.}(1998)\citenamefont {Glaser},
  \citenamefont {Schulte-Herbr{\"u}ggen}, \citenamefont {Sieveking},
  \citenamefont {Schedletzky}, \citenamefont {Nielsen}, \citenamefont
  {S{\o}rensen},\ and\ \citenamefont {Griesinger}}]{glaser1998unitary}%
  \BibitemOpen
  \bibfield  {author} {\bibinfo {author} {\bibfnamefont {S.~J.}\ \bibnamefont
  {Glaser}}, \bibinfo {author} {\bibfnamefont {T.}~\bibnamefont
  {Schulte-Herbr{\"u}ggen}}, \bibinfo {author} {\bibfnamefont {M.}~\bibnamefont
  {Sieveking}}, \bibinfo {author} {\bibfnamefont {O.}~\bibnamefont
  {Schedletzky}}, \bibinfo {author} {\bibfnamefont {N.~C.}\ \bibnamefont
  {Nielsen}}, \bibinfo {author} {\bibfnamefont {O.~W.}\ \bibnamefont
  {S{\o}rensen}}, \ and\ \bibinfo {author} {\bibfnamefont {C.}~\bibnamefont
  {Griesinger}},\ }\href@noop {} {\bibfield  {journal} {\bibinfo  {journal}
  {Science}\ }\textbf {\bibinfo {volume} {280}},\ \bibinfo {pages} {421}
  (\bibinfo {year} {1998})}\BibitemShut {NoStop}%
\bibitem [{\citenamefont {Sels}\ and\ \citenamefont
  {Polkovnikov}(2016)}]{sels2016minimizing}%
  \BibitemOpen
  \bibfield  {author} {\bibinfo {author} {\bibfnamefont {D.}~\bibnamefont
  {Sels}}\ and\ \bibinfo {author} {\bibfnamefont {A.}~\bibnamefont
  {Polkovnikov}},\ }\href@noop {} {\bibfield  {journal} {\bibinfo  {journal}
  {arXiv preprint arXiv:1607.05687}\ } (\bibinfo {year} {2016})}\BibitemShut
  {NoStop}%
\bibitem [{\citenamefont {Van~Frank}\ \emph {et~al.}(2016)\citenamefont
  {Van~Frank}, \citenamefont {Bonneau}, \citenamefont {Schmiedmayer},
  \citenamefont {Hild}, \citenamefont {Gross}, \citenamefont {Cheneau},
  \citenamefont {Bloch}, \citenamefont {Pichler}, \citenamefont {Negretti},
  \citenamefont {Calarco} \emph {et~al.}}]{van2016optimal}%
  \BibitemOpen
  \bibfield  {author} {\bibinfo {author} {\bibfnamefont {S.}~\bibnamefont
  {Van~Frank}}, \bibinfo {author} {\bibfnamefont {M.}~\bibnamefont {Bonneau}},
  \bibinfo {author} {\bibfnamefont {J.}~\bibnamefont {Schmiedmayer}}, \bibinfo
  {author} {\bibfnamefont {S.}~\bibnamefont {Hild}}, \bibinfo {author}
  {\bibfnamefont {C.}~\bibnamefont {Gross}}, \bibinfo {author} {\bibfnamefont
  {M.}~\bibnamefont {Cheneau}}, \bibinfo {author} {\bibfnamefont
  {I.}~\bibnamefont {Bloch}}, \bibinfo {author} {\bibfnamefont
  {T.}~\bibnamefont {Pichler}}, \bibinfo {author} {\bibfnamefont
  {A.}~\bibnamefont {Negretti}}, \bibinfo {author} {\bibfnamefont
  {T.}~\bibnamefont {Calarco}},  \emph {et~al.},\ }\href@noop {} {\bibfield
  {journal} {\bibinfo  {journal} {Scientific reports}\ }\textbf {\bibinfo
  {volume} {6}} (\bibinfo {year} {2016})}\BibitemShut {NoStop}%
\bibitem [{\citenamefont {Geng}\ \emph {et~al.}(2016)\citenamefont {Geng},
  \citenamefont {Wu}, \citenamefont {Wang}, \citenamefont {Xu}, \citenamefont
  {Shi}, \citenamefont {Xie}, \citenamefont {Rong},\ and\ \citenamefont
  {Du}}]{Jiangfeng2016optimalspinqubit}%
  \BibitemOpen
  \bibfield  {author} {\bibinfo {author} {\bibfnamefont {J.}~\bibnamefont
  {Geng}}, \bibinfo {author} {\bibfnamefont {Y.}~\bibnamefont {Wu}}, \bibinfo
  {author} {\bibfnamefont {X.}~\bibnamefont {Wang}}, \bibinfo {author}
  {\bibfnamefont {K.}~\bibnamefont {Xu}}, \bibinfo {author} {\bibfnamefont
  {F.}~\bibnamefont {Shi}}, \bibinfo {author} {\bibfnamefont {Y.}~\bibnamefont
  {Xie}}, \bibinfo {author} {\bibfnamefont {X.}~\bibnamefont {Rong}}, \ and\
  \bibinfo {author} {\bibfnamefont {J.}~\bibnamefont {Du}},\ }\href {\doibase
  10.1103/PhysRevLett.117.170501} {\bibfield  {journal} {\bibinfo  {journal}
  {Phys. Rev. Lett.}\ }\textbf {\bibinfo {volume} {117}},\ \bibinfo {pages}
  {170501} (\bibinfo {year} {2016})}\BibitemShut {NoStop}%
\bibitem [{\citenamefont {Agundez}\ \emph {et~al.}(2017)\citenamefont
  {Agundez}, \citenamefont {Hill}, \citenamefont {Hollenberg}, \citenamefont
  {Rogge},\ and\ \citenamefont
  {Blaauboer}}]{Superadiabaticspinchaintransfer2017}%
  \BibitemOpen
  \bibfield  {author} {\bibinfo {author} {\bibfnamefont {R.~R.}\ \bibnamefont
  {Agundez}}, \bibinfo {author} {\bibfnamefont {C.~D.}\ \bibnamefont {Hill}},
  \bibinfo {author} {\bibfnamefont {L.~C.~L.}\ \bibnamefont {Hollenberg}},
  \bibinfo {author} {\bibfnamefont {S.}~\bibnamefont {Rogge}}, \ and\ \bibinfo
  {author} {\bibfnamefont {M.}~\bibnamefont {Blaauboer}},\ }\href {\doibase
  10.1103/PhysRevA.95.012317} {\bibfield  {journal} {\bibinfo  {journal} {Phys.
  Rev. A}\ }\textbf {\bibinfo {volume} {95}},\ \bibinfo {pages} {012317}
  (\bibinfo {year} {2017})}\BibitemShut {NoStop}%
\bibitem [{\citenamefont {Baksic}\ \emph {et~al.}(2016)\citenamefont {Baksic},
  \citenamefont {Ribeiro},\ and\ \citenamefont
  {Clerk}}]{aashishcleark_diabatic}%
  \BibitemOpen
  \bibfield  {author} {\bibinfo {author} {\bibfnamefont {A.}~\bibnamefont
  {Baksic}}, \bibinfo {author} {\bibfnamefont {H.}~\bibnamefont {Ribeiro}}, \
  and\ \bibinfo {author} {\bibfnamefont {A.~A.}\ \bibnamefont {Clerk}},\ }\href
  {\doibase 10.1103/PhysRevLett.116.230503} {\bibfield  {journal} {\bibinfo
  {journal} {Phys. Rev. Lett.}\ }\textbf {\bibinfo {volume} {116}},\ \bibinfo
  {pages} {230503} (\bibinfo {year} {2016})}\BibitemShut {NoStop}%
\bibitem [{\citenamefont {Rotskoff}\ \emph {et~al.}(2017)\citenamefont
  {Rotskoff}, \citenamefont {Crooks},\ and\ \citenamefont
  {Vanden-Eijnden}}]{bulkspinoptimalcontrol}%
  \BibitemOpen
  \bibfield  {author} {\bibinfo {author} {\bibfnamefont {G.~M.}\ \bibnamefont
  {Rotskoff}}, \bibinfo {author} {\bibfnamefont {G.~E.}\ \bibnamefont
  {Crooks}}, \ and\ \bibinfo {author} {\bibfnamefont {E.}~\bibnamefont
  {Vanden-Eijnden}},\ }\href {\doibase 10.1103/PhysRevE.95.012148} {\bibfield
  {journal} {\bibinfo  {journal} {Phys. Rev. E}\ }\textbf {\bibinfo {volume}
  {95}},\ \bibinfo {pages} {012148} (\bibinfo {year} {2017})}\BibitemShut
  {NoStop}%
\bibitem [{\citenamefont {Ho}\ and\ \citenamefont
  {Hsieh}(2018)}]{ho2018efficient}%
  \BibitemOpen
  \bibfield  {author} {\bibinfo {author} {\bibfnamefont {W.~W.}\ \bibnamefont
  {Ho}}\ and\ \bibinfo {author} {\bibfnamefont {T.~H.}\ \bibnamefont {Hsieh}},\
  }\href@noop {} {\bibfield  {journal} {\bibinfo  {journal} {arXiv preprint
  arXiv:1803.00026}\ } (\bibinfo {year} {2018})}\BibitemShut {NoStop}%
\bibitem [{\citenamefont {Hegerfeldt}(2013)}]{speedlimitHegerfeldt}%
  \BibitemOpen
  \bibfield  {author} {\bibinfo {author} {\bibfnamefont {G.~C.}\ \bibnamefont
  {Hegerfeldt}},\ }\href {\doibase 10.1103/PhysRevLett.111.260501} {\bibfield
  {journal} {\bibinfo  {journal} {Phys. Rev. Lett.}\ }\textbf {\bibinfo
  {volume} {111}},\ \bibinfo {pages} {260501} (\bibinfo {year}
  {2013})}\BibitemShut {NoStop}%
\bibitem [{\citenamefont {Bao}\ \emph {et~al.}(2017)\citenamefont {Bao},
  \citenamefont {Kleer}, \citenamefont {Wang},\ and\ \citenamefont
  {Rahmani}}]{bao2017optimal}%
  \BibitemOpen
  \bibfield  {author} {\bibinfo {author} {\bibfnamefont {S.}~\bibnamefont
  {Bao}}, \bibinfo {author} {\bibfnamefont {S.}~\bibnamefont {Kleer}}, \bibinfo
  {author} {\bibfnamefont {R.}~\bibnamefont {Wang}}, \ and\ \bibinfo {author}
  {\bibfnamefont {A.}~\bibnamefont {Rahmani}},\ }\href@noop {} {\bibfield
  {journal} {\bibinfo  {journal} {arXiv preprint arXiv:1704.01423}\ } (\bibinfo
  {year} {2017})}\BibitemShut {NoStop}%
\bibitem [{\citenamefont {Yang}\ \emph {et~al.}(2017)\citenamefont {Yang},
  \citenamefont {Rahmani}, \citenamefont {Shabani}, \citenamefont {Neven},\
  and\ \citenamefont {Chamon}}]{yang2017optimizing}%
  \BibitemOpen
  \bibfield  {author} {\bibinfo {author} {\bibfnamefont {Z.-C.}\ \bibnamefont
  {Yang}}, \bibinfo {author} {\bibfnamefont {A.}~\bibnamefont {Rahmani}},
  \bibinfo {author} {\bibfnamefont {A.}~\bibnamefont {Shabani}}, \bibinfo
  {author} {\bibfnamefont {H.}~\bibnamefont {Neven}}, \ and\ \bibinfo {author}
  {\bibfnamefont {C.}~\bibnamefont {Chamon}},\ }\href@noop {} {\bibfield
  {journal} {\bibinfo  {journal} {Physical Review X}\ }\textbf {\bibinfo
  {volume} {7}},\ \bibinfo {pages} {021027} (\bibinfo {year}
  {2017})}\BibitemShut {NoStop}%
\bibitem [{\citenamefont {Bukov}\ \emph {et~al.}(2017)\citenamefont {Bukov},
  \citenamefont {Day}, \citenamefont {Sels}, \citenamefont {Weinberg},
  \citenamefont {Polkovnikov},\ and\ \citenamefont {Mehta}}]{bukov2017machine}%
  \BibitemOpen
  \bibfield  {author} {\bibinfo {author} {\bibfnamefont {M.}~\bibnamefont
  {Bukov}}, \bibinfo {author} {\bibfnamefont {A.~G.}\ \bibnamefont {Day}},
  \bibinfo {author} {\bibfnamefont {D.}~\bibnamefont {Sels}}, \bibinfo {author}
  {\bibfnamefont {P.}~\bibnamefont {Weinberg}}, \bibinfo {author}
  {\bibfnamefont {A.}~\bibnamefont {Polkovnikov}}, \ and\ \bibinfo {author}
  {\bibfnamefont {P.}~\bibnamefont {Mehta}},\ }\href@noop {} {\bibfield
  {journal} {\bibinfo  {journal} {arXiv preprint arXiv:1705.00565}\ } (\bibinfo
  {year} {2017})}\BibitemShut {NoStop}%
\bibitem [{\citenamefont {Ho}\ and\ \citenamefont
  {Zhou}(2009)}]{ho2009universal}%
  \BibitemOpen
  \bibfield  {author} {\bibinfo {author} {\bibfnamefont {T.-L.}\ \bibnamefont
  {Ho}}\ and\ \bibinfo {author} {\bibfnamefont {Q.}~\bibnamefont {Zhou}},\
  }\href@noop {} {\bibfield  {journal} {\bibinfo  {journal} {arXiv preprint
  arXiv:0911.5506}\ } (\bibinfo {year} {2009})}\BibitemShut {NoStop}%
\bibitem [{\citenamefont {Zaletel}\ \emph {et~al.}(2016)\citenamefont
  {Zaletel}, \citenamefont {Stamper-Kurn},\ and\ \citenamefont
  {Yao}}]{zaletel2016preparation}%
  \BibitemOpen
  \bibfield  {author} {\bibinfo {author} {\bibfnamefont {M.~P.}\ \bibnamefont
  {Zaletel}}, \bibinfo {author} {\bibfnamefont {D.~M.}\ \bibnamefont
  {Stamper-Kurn}}, \ and\ \bibinfo {author} {\bibfnamefont {N.~Y.}\
  \bibnamefont {Yao}},\ }\href@noop {} {\bibfield  {journal} {\bibinfo
  {journal} {arXiv preprint arXiv:1611.04591}\ } (\bibinfo {year}
  {2016})}\BibitemShut {NoStop}%
\bibitem [{\citenamefont {Agarwal}\ \emph
  {et~al.}(2017{\natexlab{b}})\citenamefont {Agarwal}, \citenamefont
  {Dalla~Torre}, \citenamefont {Schmiedmayer},\ and\ \citenamefont
  {Demler}}]{Agarwalquantumheatwaves}%
  \BibitemOpen
  \bibfield  {author} {\bibinfo {author} {\bibfnamefont {K.}~\bibnamefont
  {Agarwal}}, \bibinfo {author} {\bibfnamefont {E.~G.}\ \bibnamefont
  {Dalla~Torre}}, \bibinfo {author} {\bibfnamefont {J.}~\bibnamefont
  {Schmiedmayer}}, \ and\ \bibinfo {author} {\bibfnamefont {E.}~\bibnamefont
  {Demler}},\ }\href {\doibase 10.1103/PhysRevB.95.195157} {\bibfield
  {journal} {\bibinfo  {journal} {Phys. Rev. B}\ }\textbf {\bibinfo {volume}
  {95}},\ \bibinfo {pages} {195157} (\bibinfo {year}
  {2017}{\natexlab{b}})}\BibitemShut {NoStop}%
\bibitem [{\citenamefont {Dziarmaga}\ and\ \citenamefont
  {Rams}(2010{\natexlab{a}})}]{Dziarmagainhomogeneous}%
  \BibitemOpen
  \bibfield  {author} {\bibinfo {author} {\bibfnamefont {J.}~\bibnamefont
  {Dziarmaga}}\ and\ \bibinfo {author} {\bibfnamefont {M.~M.}\ \bibnamefont
  {Rams}},\ }\href {http://stacks.iop.org/1367-2630/12/i=5/a=055007} {\bibfield
   {journal} {\bibinfo  {journal} {New Journal of Physics}\ }\textbf {\bibinfo
  {volume} {12}},\ \bibinfo {pages} {055007} (\bibinfo {year}
  {2010}{\natexlab{a}})}\BibitemShut {NoStop}%
\bibitem [{\citenamefont {Dziarmaga}\ and\ \citenamefont
  {Rams}(2010{\natexlab{b}})}]{Dziarmagazgreaterthanone}%
  \BibitemOpen
  \bibfield  {author} {\bibinfo {author} {\bibfnamefont {J.}~\bibnamefont
  {Dziarmaga}}\ and\ \bibinfo {author} {\bibfnamefont {M.~M.}\ \bibnamefont
  {Rams}},\ }\href {http://stacks.iop.org/1367-2630/12/i=10/a=103002}
  {\bibfield  {journal} {\bibinfo  {journal} {New Journal of Physics}\ }\textbf
  {\bibinfo {volume} {12}},\ \bibinfo {pages} {103002} (\bibinfo {year}
  {2010}{\natexlab{b}})}\BibitemShut {NoStop}%
\bibitem [{\citenamefont {Agarwal}\ \emph {et~al.}(2014)\citenamefont
  {Agarwal}, \citenamefont {Torre}, \citenamefont {Rauer}, \citenamefont
  {Langen}, \citenamefont {Schmiedmayer},\ and\ \citenamefont
  {Demler}}]{AgarwalChiral}%
  \BibitemOpen
  \bibfield  {author} {\bibinfo {author} {\bibfnamefont {K.}~\bibnamefont
  {Agarwal}}, \bibinfo {author} {\bibfnamefont {E.~G.~D.}\ \bibnamefont
  {Torre}}, \bibinfo {author} {\bibfnamefont {B.}~\bibnamefont {Rauer}},
  \bibinfo {author} {\bibfnamefont {T.}~\bibnamefont {Langen}}, \bibinfo
  {author} {\bibfnamefont {J.}~\bibnamefont {Schmiedmayer}}, \ and\ \bibinfo
  {author} {\bibfnamefont {E.}~\bibnamefont {Demler}},\ }\href {\doibase
  10.1103/PhysRevLett.113.190401} {\bibfield  {journal} {\bibinfo  {journal}
  {Phys. Rev. Lett.}\ }\textbf {\bibinfo {volume} {113}},\ \bibinfo {pages}
  {190401} (\bibinfo {year} {2014})}\BibitemShut {NoStop}%
\bibitem [{\citenamefont {Bhaseen}\ \emph {et~al.}(2015)\citenamefont
  {Bhaseen}, \citenamefont {Doyon}, \citenamefont {Lucas},\ and\ \citenamefont
  {Schalm}}]{bhaseen2015energy}%
  \BibitemOpen
  \bibfield  {author} {\bibinfo {author} {\bibfnamefont {M.}~\bibnamefont
  {Bhaseen}}, \bibinfo {author} {\bibfnamefont {B.}~\bibnamefont {Doyon}},
  \bibinfo {author} {\bibfnamefont {A.}~\bibnamefont {Lucas}}, \ and\ \bibinfo
  {author} {\bibfnamefont {K.}~\bibnamefont {Schalm}},\ }\href@noop {}
  {\bibfield  {journal} {\bibinfo  {journal} {Nature Physics}\ }\textbf
  {\bibinfo {volume} {11}},\ \bibinfo {pages} {509} (\bibinfo {year}
  {2015})}\BibitemShut {NoStop}%
\bibitem [{\citenamefont {Bernard}\ and\ \citenamefont
  {Doyon}(2012)}]{bernard2012energy}%
  \BibitemOpen
  \bibfield  {author} {\bibinfo {author} {\bibfnamefont {D.}~\bibnamefont
  {Bernard}}\ and\ \bibinfo {author} {\bibfnamefont {B.}~\bibnamefont
  {Doyon}},\ }\href@noop {} {\bibfield  {journal} {\bibinfo  {journal} {Journal
  of Physics A: Mathematical and Theoretical}\ }\textbf {\bibinfo {volume}
  {45}},\ \bibinfo {pages} {362001} (\bibinfo {year} {2012})}\BibitemShut
  {NoStop}%
\bibitem [{\citenamefont {Castro-Alvaredo}\ \emph {et~al.}(2016)\citenamefont
  {Castro-Alvaredo}, \citenamefont {Doyon},\ and\ \citenamefont
  {Yoshimura}}]{castro2016emergent}%
  \BibitemOpen
  \bibfield  {author} {\bibinfo {author} {\bibfnamefont {O.~A.}\ \bibnamefont
  {Castro-Alvaredo}}, \bibinfo {author} {\bibfnamefont {B.}~\bibnamefont
  {Doyon}}, \ and\ \bibinfo {author} {\bibfnamefont {T.}~\bibnamefont
  {Yoshimura}},\ }\href@noop {} {\bibfield  {journal} {\bibinfo  {journal}
  {Physical Review X}\ }\textbf {\bibinfo {volume} {6}},\ \bibinfo {pages}
  {041065} (\bibinfo {year} {2016})}\BibitemShut {NoStop}%
\bibitem [{\citenamefont {Lucas}\ \emph {et~al.}(2016)\citenamefont {Lucas},
  \citenamefont {Schalm}, \citenamefont {Doyon},\ and\ \citenamefont
  {Bhaseen}}]{lucas2016shock}%
  \BibitemOpen
  \bibfield  {author} {\bibinfo {author} {\bibfnamefont {A.}~\bibnamefont
  {Lucas}}, \bibinfo {author} {\bibfnamefont {K.}~\bibnamefont {Schalm}},
  \bibinfo {author} {\bibfnamefont {B.}~\bibnamefont {Doyon}}, \ and\ \bibinfo
  {author} {\bibfnamefont {M.}~\bibnamefont {Bhaseen}},\ }\href@noop {}
  {\bibfield  {journal} {\bibinfo  {journal} {Physical Review D}\ }\textbf
  {\bibinfo {volume} {94}},\ \bibinfo {pages} {025004} (\bibinfo {year}
  {2016})}\BibitemShut {NoStop}%
\bibitem [{\citenamefont {Fulling}\ \emph {et~al.}(1974)\citenamefont
  {Fulling}, \citenamefont {Parker},\ and\ \citenamefont
  {Hu}}]{Fullingconformalvacuum}%
  \BibitemOpen
  \bibfield  {author} {\bibinfo {author} {\bibfnamefont {S.~A.}\ \bibnamefont
  {Fulling}}, \bibinfo {author} {\bibfnamefont {L.}~\bibnamefont {Parker}}, \
  and\ \bibinfo {author} {\bibfnamefont {B.~L.}\ \bibnamefont {Hu}},\ }\href
  {\doibase 10.1103/PhysRevD.10.3905} {\bibfield  {journal} {\bibinfo
  {journal} {Phys. Rev. D}\ }\textbf {\bibinfo {volume} {10}},\ \bibinfo
  {pages} {3905} (\bibinfo {year} {1974})}\BibitemShut {NoStop}%
\bibitem [{\citenamefont {Chandran}\ \emph {et~al.}(2013)\citenamefont
  {Chandran}, \citenamefont {Nanduri}, \citenamefont {Gubser},\ and\
  \citenamefont {Sondhi}}]{chandrancoarsening}%
  \BibitemOpen
  \bibfield  {author} {\bibinfo {author} {\bibfnamefont {A.}~\bibnamefont
  {Chandran}}, \bibinfo {author} {\bibfnamefont {A.}~\bibnamefont {Nanduri}},
  \bibinfo {author} {\bibfnamefont {S.~S.}\ \bibnamefont {Gubser}}, \ and\
  \bibinfo {author} {\bibfnamefont {S.~L.}\ \bibnamefont {Sondhi}},\ }\href
  {\doibase 10.1103/PhysRevB.88.024306} {\bibfield  {journal} {\bibinfo
  {journal} {Phys. Rev. B}\ }\textbf {\bibinfo {volume} {88}},\ \bibinfo
  {pages} {024306} (\bibinfo {year} {2013})}\BibitemShut {NoStop}%
\bibitem [{\citenamefont {Smacchia}\ \emph {et~al.}(2015)\citenamefont
  {Smacchia}, \citenamefont {Knap}, \citenamefont {Demler},\ and\ \citenamefont
  {Silva}}]{Smacchia2015dQPTinlargeNON}%
  \BibitemOpen
  \bibfield  {author} {\bibinfo {author} {\bibfnamefont {P.}~\bibnamefont
  {Smacchia}}, \bibinfo {author} {\bibfnamefont {M.}~\bibnamefont {Knap}},
  \bibinfo {author} {\bibfnamefont {E.}~\bibnamefont {Demler}}, \ and\ \bibinfo
  {author} {\bibfnamefont {A.}~\bibnamefont {Silva}},\ }\href {\doibase
  10.1103/PhysRevB.91.205136} {\bibfield  {journal} {\bibinfo  {journal} {Phys.
  Rev. B}\ }\textbf {\bibinfo {volume} {91}},\ \bibinfo {pages} {205136}
  (\bibinfo {year} {2015})}\BibitemShut {NoStop}%
\bibitem [{\citenamefont {Unruh}(1976)}]{Unruhoriginal}%
  \BibitemOpen
  \bibfield  {author} {\bibinfo {author} {\bibfnamefont {W.~G.}\ \bibnamefont
  {Unruh}},\ }\href {\doibase 10.1103/PhysRevD.14.870} {\bibfield  {journal}
  {\bibinfo  {journal} {Phys. Rev. D}\ }\textbf {\bibinfo {volume} {14}},\
  \bibinfo {pages} {870} (\bibinfo {year} {1976})}\BibitemShut {NoStop}%
\bibitem [{\citenamefont {Crispino}\ \emph {et~al.}(2008)\citenamefont
  {Crispino}, \citenamefont {Higuchi},\ and\ \citenamefont
  {Matsas}}]{unruheffectandapplications}%
  \BibitemOpen
  \bibfield  {author} {\bibinfo {author} {\bibfnamefont {L.~C.~B.}\
  \bibnamefont {Crispino}}, \bibinfo {author} {\bibfnamefont {A.}~\bibnamefont
  {Higuchi}}, \ and\ \bibinfo {author} {\bibfnamefont {G.~E.~A.}\ \bibnamefont
  {Matsas}},\ }\href {\doibase 10.1103/RevModPhys.80.787} {\bibfield  {journal}
  {\bibinfo  {journal} {Rev. Mod. Phys.}\ }\textbf {\bibinfo {volume} {80}},\
  \bibinfo {pages} {787} (\bibinfo {year} {2008})}\BibitemShut {NoStop}%
\bibitem [{\citenamefont {Gaunt}\ \emph {et~al.}(2013)\citenamefont {Gaunt},
  \citenamefont {Schmidutz}, \citenamefont {Gotlibovych}, \citenamefont
  {Smith},\ and\ \citenamefont {Hadzibabic}}]{HadzibabicUniformBose}%
  \BibitemOpen
  \bibfield  {author} {\bibinfo {author} {\bibfnamefont {A.~L.}\ \bibnamefont
  {Gaunt}}, \bibinfo {author} {\bibfnamefont {T.~F.}\ \bibnamefont
  {Schmidutz}}, \bibinfo {author} {\bibfnamefont {I.}~\bibnamefont
  {Gotlibovych}}, \bibinfo {author} {\bibfnamefont {R.~P.}\ \bibnamefont
  {Smith}}, \ and\ \bibinfo {author} {\bibfnamefont {Z.}~\bibnamefont
  {Hadzibabic}},\ }\href {\doibase 10.1103/PhysRevLett.110.200406} {\bibfield
  {journal} {\bibinfo  {journal} {Phys. Rev. Lett.}\ }\textbf {\bibinfo
  {volume} {110}},\ \bibinfo {pages} {200406} (\bibinfo {year}
  {2013})}\BibitemShut {NoStop}%
\bibitem [{\citenamefont {Mukherjee}\ \emph {et~al.}(2015)\citenamefont
  {Mukherjee}, \citenamefont {Ku}, \citenamefont {Yan}, \citenamefont {Patel},
  \citenamefont {Guardado-Sanchez}, \citenamefont {Yefsah}, \citenamefont
  {Struck}, \citenamefont {Zwierlein} \emph {et~al.}}]{mukherjee2015fermi}%
  \BibitemOpen
  \bibfield  {author} {\bibinfo {author} {\bibfnamefont {B.}~\bibnamefont
  {Mukherjee}}, \bibinfo {author} {\bibfnamefont {M.}~\bibnamefont {Ku}},
  \bibinfo {author} {\bibfnamefont {Z.}~\bibnamefont {Yan}}, \bibinfo {author}
  {\bibfnamefont {P.}~\bibnamefont {Patel}}, \bibinfo {author} {\bibfnamefont
  {E.}~\bibnamefont {Guardado-Sanchez}}, \bibinfo {author} {\bibfnamefont
  {T.}~\bibnamefont {Yefsah}}, \bibinfo {author} {\bibfnamefont
  {J.}~\bibnamefont {Struck}}, \bibinfo {author} {\bibfnamefont
  {M.}~\bibnamefont {Zwierlein}},  \emph {et~al.},\ }in\ \href@noop {} {\emph
  {\bibinfo {booktitle} {APS Division of Atomic, Molecular and Optical Physics
  Meeting Abstracts}}},\ Vol.~\bibinfo {volume} {1}\ (\bibinfo {year} {2015})\
  p.\ \bibinfo {pages} {7008}\BibitemShut {NoStop}%
\bibitem [{\citenamefont {Ergül}\ \emph {et~al.}(2013)\citenamefont {Ergül},
  \citenamefont {Lidmar}, \citenamefont {Johansson}, \citenamefont {Azizoğlu},
  \citenamefont {Schaeffer},\ and\ \citenamefont {Haviland}}]{HavilandArray}%
  \BibitemOpen
  \bibfield  {author} {\bibinfo {author} {\bibfnamefont {A.}~\bibnamefont
  {Ergül}}, \bibinfo {author} {\bibfnamefont {J.}~\bibnamefont {Lidmar}},
  \bibinfo {author} {\bibfnamefont {J.}~\bibnamefont {Johansson}}, \bibinfo
  {author} {\bibfnamefont {Y.}~\bibnamefont {Azizoğlu}}, \bibinfo {author}
  {\bibfnamefont {D.}~\bibnamefont {Schaeffer}}, \ and\ \bibinfo {author}
  {\bibfnamefont {D.~B.}\ \bibnamefont {Haviland}},\ }\href
  {http://stacks.iop.org/1367-2630/15/i=9/a=095014} {\bibfield  {journal}
  {\bibinfo  {journal} {New Journal of Physics}\ }\textbf {\bibinfo {volume}
  {15}},\ \bibinfo {pages} {095014} (\bibinfo {year} {2013})}\BibitemShut
  {NoStop}%
\bibitem [{\citenamefont {Zimmer}\ \emph {et~al.}(2013)\citenamefont {Zimmer},
  \citenamefont {Vogt}, \citenamefont {Fiebig}, \citenamefont {Syzranov},
  \citenamefont {Lukashenko}, \citenamefont {Sch\"afer}, \citenamefont
  {Rotzinger}, \citenamefont {Shnirman}, \citenamefont {Marthaler},\ and\
  \citenamefont {Ustinov}}]{UstinovJJArray}%
  \BibitemOpen
  \bibfield  {author} {\bibinfo {author} {\bibfnamefont {J.}~\bibnamefont
  {Zimmer}}, \bibinfo {author} {\bibfnamefont {N.}~\bibnamefont {Vogt}},
  \bibinfo {author} {\bibfnamefont {A.}~\bibnamefont {Fiebig}}, \bibinfo
  {author} {\bibfnamefont {S.~V.}\ \bibnamefont {Syzranov}}, \bibinfo {author}
  {\bibfnamefont {A.}~\bibnamefont {Lukashenko}}, \bibinfo {author}
  {\bibfnamefont {R.}~\bibnamefont {Sch\"afer}}, \bibinfo {author}
  {\bibfnamefont {H.}~\bibnamefont {Rotzinger}}, \bibinfo {author}
  {\bibfnamefont {A.}~\bibnamefont {Shnirman}}, \bibinfo {author}
  {\bibfnamefont {M.}~\bibnamefont {Marthaler}}, \ and\ \bibinfo {author}
  {\bibfnamefont {A.~V.}\ \bibnamefont {Ustinov}},\ }\href {\doibase
  10.1103/PhysRevB.88.144506} {\bibfield  {journal} {\bibinfo  {journal} {Phys.
  Rev. B}\ }\textbf {\bibinfo {volume} {88}},\ \bibinfo {pages} {144506}
  (\bibinfo {year} {2013})}\BibitemShut {NoStop}%
\bibitem [{\citenamefont {Blatt}\ and\ \citenamefont
  {Roos}(2012)}]{blatt2012quantumsimulation}%
  \BibitemOpen
  \bibfield  {author} {\bibinfo {author} {\bibfnamefont {R.}~\bibnamefont
  {Blatt}}\ and\ \bibinfo {author} {\bibfnamefont {C.}~\bibnamefont {Roos}},\
  }\href@noop {} {\bibfield  {journal} {\bibinfo  {journal} {Nature Physics}\
  }\textbf {\bibinfo {volume} {8}},\ \bibinfo {pages} {277} (\bibinfo {year}
  {2012})}\BibitemShut {NoStop}%
\bibitem [{\citenamefont {Smith}\ \emph {et~al.}(2016)\citenamefont {Smith},
  \citenamefont {Lee}, \citenamefont {Richerme}, \citenamefont {Neyenhuis},
  \citenamefont {Hess}, \citenamefont {Hauke}, \citenamefont {Heyl},
  \citenamefont {Huse},\ and\ \citenamefont {Monroe}}]{MBLTrappedIonMonroe}%
  \BibitemOpen
  \bibfield  {author} {\bibinfo {author} {\bibfnamefont {J.}~\bibnamefont
  {Smith}}, \bibinfo {author} {\bibfnamefont {A.}~\bibnamefont {Lee}}, \bibinfo
  {author} {\bibfnamefont {P.}~\bibnamefont {Richerme}}, \bibinfo {author}
  {\bibfnamefont {B.}~\bibnamefont {Neyenhuis}}, \bibinfo {author}
  {\bibfnamefont {P.~W.}\ \bibnamefont {Hess}}, \bibinfo {author}
  {\bibfnamefont {P.}~\bibnamefont {Hauke}}, \bibinfo {author} {\bibfnamefont
  {M.}~\bibnamefont {Heyl}}, \bibinfo {author} {\bibfnamefont {D.~A.}\
  \bibnamefont {Huse}}, \ and\ \bibinfo {author} {\bibfnamefont
  {C.}~\bibnamefont {Monroe}},\ }\href {\doibase
  http://dx.doi.org/10.1038/nphys3783} {\bibfield  {journal} {\bibinfo
  {journal} {Nature Physics}\ } (\bibinfo {year} {2016}),\
  http://dx.doi.org/10.1038/nphys3783}\BibitemShut {NoStop}%
\bibitem [{\citenamefont {Rajabi}\ \emph {et~al.}(2018)\citenamefont {Rajabi},
  \citenamefont {Motlakunta}, \citenamefont {Shih}, \citenamefont
  {Kotibhaskar}, \citenamefont {Quraishi}, \citenamefont {Ajoy},\ and\
  \citenamefont {Islam}}]{rajabi2018dynamic}%
  \BibitemOpen
  \bibfield  {author} {\bibinfo {author} {\bibfnamefont {F.}~\bibnamefont
  {Rajabi}}, \bibinfo {author} {\bibfnamefont {S.}~\bibnamefont {Motlakunta}},
  \bibinfo {author} {\bibfnamefont {C.-Y.}\ \bibnamefont {Shih}}, \bibinfo
  {author} {\bibfnamefont {N.}~\bibnamefont {Kotibhaskar}}, \bibinfo {author}
  {\bibfnamefont {Q.}~\bibnamefont {Quraishi}}, \bibinfo {author}
  {\bibfnamefont {A.}~\bibnamefont {Ajoy}}, \ and\ \bibinfo {author}
  {\bibfnamefont {R.}~\bibnamefont {Islam}},\ }\href@noop {} {\bibfield
  {journal} {\bibinfo  {journal} {arXiv preprint arXiv:1808.06124}\ } (\bibinfo
  {year} {2018})}\BibitemShut {NoStop}%
\bibitem [{\citenamefont {Gring}\ \emph {et~al.}(2012)\citenamefont {Gring},
  \citenamefont {Kuhnert}, \citenamefont {Langen}, \citenamefont {Kitagawa},
  \citenamefont {Rauer}, \citenamefont {Schreitl}, \citenamefont {Mazets},
  \citenamefont {Smith}, \citenamefont {Demler},\ and\ \citenamefont
  {Schmiedmayer}}]{Gring}%
  \BibitemOpen
  \bibfield  {author} {\bibinfo {author} {\bibfnamefont {M.}~\bibnamefont
  {Gring}}, \bibinfo {author} {\bibfnamefont {M.}~\bibnamefont {Kuhnert}},
  \bibinfo {author} {\bibfnamefont {T.}~\bibnamefont {Langen}}, \bibinfo
  {author} {\bibfnamefont {T.}~\bibnamefont {Kitagawa}}, \bibinfo {author}
  {\bibfnamefont {B.}~\bibnamefont {Rauer}}, \bibinfo {author} {\bibfnamefont
  {M.}~\bibnamefont {Schreitl}}, \bibinfo {author} {\bibfnamefont
  {I.}~\bibnamefont {Mazets}}, \bibinfo {author} {\bibfnamefont {D.~A.}\
  \bibnamefont {Smith}}, \bibinfo {author} {\bibfnamefont {E.}~\bibnamefont
  {Demler}}, \ and\ \bibinfo {author} {\bibfnamefont {J.}~\bibnamefont
  {Schmiedmayer}},\ }\href@noop {} {\bibfield  {journal} {\bibinfo  {journal}
  {Science}\ }\textbf {\bibinfo {volume} {337}},\ \bibinfo {pages} {1318}
  (\bibinfo {year} {2012})}\BibitemShut {NoStop}%
\end{thebibliography}
%

\end{document}